\documentclass[conference]{IEEEtran}
\IEEEoverridecommandlockouts
\usepackage{cite}
\usepackage{amsmath,amssymb,amsfonts}
\usepackage{algorithmic}
\usepackage{graphicx}
\usepackage{textcomp}
\usepackage{xcolor}
\usepackage{algorithm}
\usepackage{bbm}
\usepackage{cuted}
\usepackage{subfigure}

\newcommand{\mcone}{\textcolor{blue}}

\usepackage{hyperref}
\def\BibTeX{{\rm B\kern-.05em{\sc i\kern-.025em b}\kern-.08em
		T\kern-.1667em\lower.7ex\hbox{E}\kern-.125emX}}

\begin{document}
	
	\title{Fluid-Antenna Enhanced ISAC: Joint Antenna Positioning and Dual-Functional Beamforming Design under Perfect and Imperfect CSI}
	
	\author{\IEEEauthorblockN{Tian Hao, Changxin Shi, Qingqing Wu, Bin Xia, Yinghong Guo, Lianghui Ding, and Feng Yang} 	\thanks{Tian Hao, Changxin Shi, Qingqing Wu, Bin Xia, Yinghong Guo, Lianghui Ding, and Feng Yang are with the School of Electronic Information and Electrical Engineering, SJTU, Shanghai 200240, China. An earlier version of this paper was presented in part at IEEE/CIC ICCC 2024 \cite{ICCC_H_2024}.}}

	\maketitle
	
	\begin{abstract}
		

	Integrated sensing and communication (ISAC) emerges as an essential technique for overcoming spectrum congestion. However, the performance of traditional ISAC systems with fixed-position-antennas (FPA) is limited due to insufficient spatial degree of freedom (DoF) exploration. Recently, fluid antenna (FA) \mcone{with reconfigurable antenna position} is developed to enhance the sensing and communication performance by reshaping the channel. This paper investigates an FA-enhanced ISAC system where a base station is equipped with multiple FAs to \mcone{communicate} with multiple single-antenna users and with FPAs to sense a point target. In this paper, we consider \mcone{both} perfect and imperfect channel state information (CSI) of the communication channel and sensing channel. In two cases, we focus on the maximization of the sensing signal-to-noise (SNR) by optimizing the positions of FAs and the dual-functional beamforming under the constraints of the FA moving region, the minimum FA distance and the minimum signal-to-interference-plus-noise (SINR) per user. Specifically, for the ideal case of perfect CSI, an iterative alternating optimization (AO) algorithm is proposed to tackle the formulated problem where the dual-functional beamforming and the FA positions are obtained via semidefinite relaxation (SDR) and successive convex approximation (SCA) techniques. Then, for the imperfect CSI case, we propose an AO-based iterative algorithm where $\mathcal{S}-$Procedure and SCA are applied to obtain the dual-functional beamforming and the FA positions. Furthermore, we analytically and numerically prove the convergence of the proposed algorithms. Numerical results demonstrate the notable gains of the proposed algorithms in the respective cases.

	\end{abstract}
	
	
	\section{Introduction}
	With the explosive growth in the number of communication and sensing devices, wireless networks demand a considerable capacity and a high throughput, which causes an increasing pressure on spectrum resources. As a promising solution, integrated sensing and communication (ISAC)\cite{b1} or dual-functional radar-communication \cite{b3} has been recently proposed to meet the requirements by deploying both communication and sensing tasks on a hardware platform and sharing the radar spectrum with the communication system. By carefully designing the transceiver hardware architectures and the signal processing methods \cite{b2,b3,b4,b5}, \mcone{an} ISAC system is superior to \mcone{an} individual system (radar system or communication system) in spectral efficiency and hardware cost. Therefore, the ISAC system has attracted much interest from both the industry and academic community.
	
	According to whether perfect channel information state (CSI) is available, the existing optimization-based ISAC works can be classified into two cases: perfect CSI case \cite{M1,M2,M3,M4} and imperfect CSI case \cite{RM1,RM2,RM3}. In both cases, \mcone{an} ISAC system consists of \mcone{a} base station (BS), \mcone{multiple} communication users and \mcone{a} sensing target, and the BS typically equips with transmit antennas for dual-functional waveform transmission and receive antennas for sensing reception. Specifically, in \cite{M1,M2}, the authors optimize the dual-functional beamforming to maximize the sensing performance, i.e., beampattern matching mean square error (MSE) \cite{M1} and radar signal-to-interference-plus-noise (SINR) \cite{M2}, while ensuring the quality-of-service (QoS), i.e., SINR requirement\cite{M1,M2}, of each communication user in a \mcone{multi-user multiple-input multiple-output} (MIMO) ISAC system. Furthermore, the authors of \cite{M3,M4} optimize the dual-functional beamforming to maximize the communication performance, i.e., communication rate \cite{M3} and communication spectral efficiency \cite{M4}, while ensuring the sensing performance, i.e., sensing signal-to-noise (SNR) \cite{M3} and sensing beampattern gain \cite{M4}, in a multi-user MIMO ISAC system. Although satisfactory results have been achieved in the above works, the works above depend on the perfect CSI in the system. The designs of the above works may not work well and lead to some performance loss under imperfect CSI. Thus, some initial works study the beamforming design for the ISAC system under imperfect CSI assumption \cite{RM1,RM2,RM3}. Specifically, in \cite{RM1}, the authors optimize the dual-functional beamforming under the imperfect CSI of the communication channel to minimize the beampattern matching MSE while guaranteeing the minimum communication rate in an ISAC system. Moreover, the authors of \cite{RM2,RM3} optimize the dual-functional beamforming design in the ISAC system under the imperfect CSI of the communication channel and sensing channel to maximize the sensing SNR while ensuring the minimum SINR requirement of each communication user. However, due to the fixed position of antennas (FPA) in these ISAC systems \cite{M1,M2,M3,M4,RM1,RM2,RM3}, the diversity and spatial multiplexing gains are limited since the full spatial variation of the wireless channel (different positions of each antenna) in a given spatial field cannot be explored. 
		
	Recently, fluid antenna (FA) \cite{b6}, also named movable antenna \cite{M6}, has been developed as a promising technique to exploit the wireless channel variation of each antenna in a continuous spatial domain \cite{b6,b7,M6,b9}. Specifically, in FA-enabled systems, each FA is connected to a single RF chain via a flexible cable, and the position of FA can be adjusted in a one-dimensional (1D) region \cite{b6}, two-dimensional (2D) region \cite{b7} \mcone{and three-dimensional (3D) space \cite{b9}}. Accordingly, the wireless channel can be configured to improve the system performance by exploring the CSI \mcone{function} of different locations of each FA \cite{M6}.  
	
	Existing works propose various FA-based optimization schemes to verify the outstanding performance of the FA-assisted system compared with the traditional FPA-based system \cite{b9,b901,b101,b102,b103,b104,b11}. \mcone{First, some works exploit the FA-based uplink communication scenario \cite{b9,b901}}. In a wireless uplink communication system with a multi-FPA BS and multiple single-FA users \cite{b9}, the authors minimize the total transmit power by jointly optimizing the antenna positions \mcone{in 3D space} and the receive beamforming. In an uplink communication system with a multi-FA BS and multiple single-FPA users \cite{b901}, the authors minimize the transmit power by optimizing the FA positions \mcone{in 1D region}, while guaranteeing the minimum communication rate. \mcone{Then, there are also some works exploit the FA-based downlink communication scenario. \cite{b101,b102,b103,b104,b11}}. In a wireless downlink communication system with a multi-FA BS together with single-FPA users \cite{b101} or single-FA users \cite{b102}, the authors of \cite{b101,b102} maximize the communication rate by jointly optimizing the beamforming and antenna positions \mcone{in 2D region}, while ensuring the power constraint. In a downlink communication system with a multi-FPA BS and multiple single-FA users, the authors of \cite{b103} minimize the total transmit power by jointly optimizing the antenna positions \mcone{in 2D region} and the transmit beamforming. In a multiple-input single-output system with several multi-FAs BSs and single-FPA users, the authors of \cite{b104} minimize the total transmit power and inter-user interference by jointly optimizing the FA positions \mcone{in 2D region} and transmit beamforming. In a MIMO system with a multi-FAs BS and a muli-FAs user \cite{b11}, the authors maximize the channel capacity by jointly optimizing the transmit beamforming and the positions of the transmit FAs and the receive FAs \mcone{in 2D region}, while guaranteeing the minimum distance between FAs. For an individual communication system, FA has a significant ability to improve communication performance \cite{b9,b901,b101,b102,b103,b104,b11}. However, these designs are based on perfect CSI, which may not be suitable for the imperfect CSI case. These inspire us to investigate the system performance by applying FAs in the ISAC system under both perfect and imperfect CSI cases. 
	
	This paper studies the FA-enhanced ISAC system where the BS is equipped with FAs as transmitting antennas and with FPAs as receiving antennas, serves multiple communication users while simultaneously sensing one point target. The FAs not only enhance the downlink communication from the BS to users, but also exploit more spatial degree of freedom (DoF) for sensing. In this paper, we consider two cases of whether perfect CSI of the communication channel and sensing channel are available at the BS. Throughout the paper, we aim to maximize the sensing SNR by jointly optimizing the positions of FAs and dual-functional beamforming at the BS, subject to the constraints of the finite moving region of FAs, the minimum FA distance, and the minimum SINR requirement of each communication user. The main contributions of this paper are summarized in the following:
	\begin{itemize}
		\item[$\bullet$] First, we consider an ideal case of perfect CSI. The resulting problem is highly challenging due to the coupled optimization variables in both constraints and objective function. To tackle the formulated problem, we propose an alternating optimization (AO) based iterative algorithm where in each iteration, the dual-functional beamforming and the FA positions are obtained via semidefinite relaxation (SDR) and successive convex approximation (SCA), respectively. Then, we prove the convergence of the proposed algorithm. \mcone{The study in this case serves as a performance upper bound of the practical case with non-ideal CSI and helps evaluate the corresponding designs.}
		\item[$\bullet$] Next, we consider a practical scenario with imperfect CSI of the communication channel and the sensing channel. The resulting problem is more complicated than the former one due to an infinite number of constraints. To solve this intractable problem, we propose a new AO-based iterative algorithm in which the the $\mathcal{S}-$Procedure and SCA techniques are applied in each iteration to obtain the dual-functional beamforming and the FA positions. Note that the convergence of the proposed algorithm is numerically shown in the simulation.
		\item[$\bullet$] Finally, we numerically validate the performance of the proposed designs under perfect and imperfect CSI cases. Simulation results show that the proposed designs can significantly improve the sensing SNR up to $8.72\%-178.85\%$ compared with existing baselines, indicating the superiority of the proposed designs.
	\end{itemize}
	
	The rest of this paper is organized as follows. Section $\text{\uppercase\expandafter{\romannumeral 2}}$ introduces the system model and formulates two problems. In Section $\text{\uppercase\expandafter{\romannumeral 3}}$ and Section $\text{\uppercase\expandafter{\romannumeral 4}}$, two different AO-based iterative algorithms are proposed to solve the representative problems. Section $\text{\uppercase\expandafter{\romannumeral 5}}$ presents numerical results and Section $\text{\uppercase\expandafter{\romannumeral 6}}$ concludes the paper.
	
	$\textit{Notations:}$ In this paper, $\left( \cdot \right)^T$, $\left( \cdot \right)^*$, and $\left( \cdot \right)^H$ denote the operations of transpose, conjugate, and Hermitian transpose. Matrices and vectors are denoted by boldface upper-case and lower-case letters. $\mathbb{C}^{N\times M}$ and $\mathbb{R}^{N\times M}$ denote the space of $N \times M$ matrices with complex and real entries. $\text{diag}\left\{\mathbf{a}\right\}$ denotes the diagonal matrix whose diagonal entries are composed of vector $\mathbf{a}$. The $\left( \textit{i,j} \right)$-th element of matrix $\mathbf{A}$ is denoted by $\mathbf{A}(i,j)$. $\Vert \cdot \Vert_2$ and $\Vert \cdot \Vert_F$ denote the $\textit{l}_2$-norm and Frobenius norm, respectively. $|a|$ and $\angle a$ denote the amplitude and phase of complex number $a$. $\text{Tr}(\mathbf{X})$ and $\text{rank}(\mathbf{X})$ denote the trace and rank of $\mathbf{X}$, respectively.
	
	\section{System Model and Problem Formulation}
	\subsection{FA-Enabled ISAC System}

	We consider an ISAC system where the BS interacts with $K$ communication users (CUs), each of which is equipped with a single fixed position antenna (FPA) while sensing a point target by a dual-functional signal. For convenience, let $\mathcal{K}\triangleq \left\{1,2,...,K \right\}$ denote the set of user indices. The BS is equipped with a planer array with $N_t$ fluid antennas (FAs) for signal transmission and a uniform rectangular array consisting of $P \times Q$ FPAs for signal reception \footnote[1]{Due to the constraints of algorithm complexity and hardware cost, the antennas for signal reception are assumed to be fixed.}. The channel condition can be configured by changing the FAs' position in a rectangular region $\mathcal{C}_t$ of size $W\times L$ ($m \times m$) \cite{M6}, the center of which is denoted as the origin point of $\mathcal{C}_t$. The position of the $m$-th FA is represented by its coordinates on $\mathcal{C}_t$, i.e.,  $\mathbf{t}_m\triangleq[x_m,y_m]^T\in\mathcal{C}_t$.
	
	\begin{figure}[htbp]
		\centering{\includegraphics[width=1\linewidth]{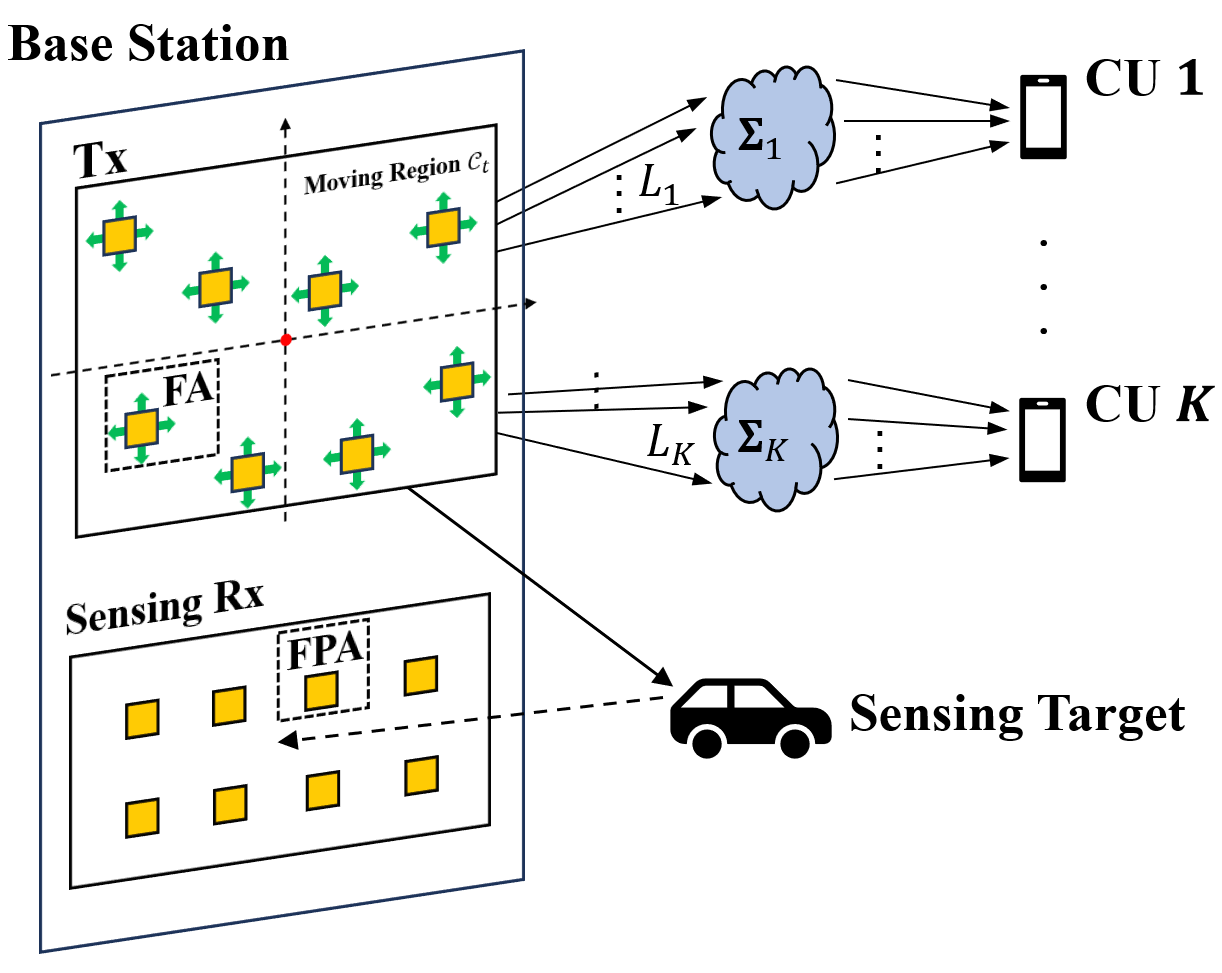}}
		\caption{ An FA-aided ISAC system.}
		\label{systemmodel}
	\end{figure} 
	
	Let $\mathbf{w}_k\in \mathbb{C}^{N_t\times 1}$ represent the beamforming for user $k$. The signal $\mathbf{x}\in \mathbb{C}^{N_t\times 1}$ transmitted by ISAC BS can be written as:
	\begin{equation} 
		\mathbf{x}\triangleq\sum_{k=1}^{K}\mathbf{w}_k s_k=\mathbf{W}\mathbf{s},
	\end{equation}
	where $\mathbf{s}=[s_1,...,s_K]^T\in \mathbb{C}^{K\times 1}$  with $\mathbb{E}[\mathbf{s}\mathbf{s}^H]=\mathbf{I}_K$ represents the data symbols for the CUs, and $\mathbf{W}=[\mathbf{w}_1,...,\mathbf{w}_K]\in \mathbb{C}^{N_t\times K}$. Note that $\mathbf{x}$ is the dual-functional signal, which can be used for both sensing and communication \cite{b5}, and $\mathbf{W}$ is a dual-functional beamforming. Then, we have the following constraint for the transmit power of the BS:
	\begin{equation}
		E(\Vert \mathbf{x} \Vert_2^2) \triangleq \text{Tr}(\mathbf{W}\mathbf{W}^H)\leq P_{max},\label{power_cons}
	\end{equation}
	where $P_{max}$ represents the maximum available power budget at the BS.
	
	\subsection{Communication Model}
	We adopt a far field-response based channel model for communication channel. Since the sizes of moving region for the FAs are much smaller than the signal propagation distance of each path between BS and CU, the angle-of-arrival (AoA), the angle-of-departure (AoD), and the amplitude of path response remain constant at different positions in FAs \cite{M6}. Let $L_k$ denote the number of paths between the BS and the user $k$. For user $k$, the signal propagation difference for the $l$-th channel path between the position of $m$-th FA and the origin point of the moving region is written as 
	\begin{align}
		&\rho(\mathbf{t}_m,\theta_{k,l},\phi_{k,l})=x_m \cos\theta_{k,l}\sin\phi_{k,l} + y_m\sin\theta_{k,l}, \nonumber \\
		&\qquad \qquad \qquad \qquad \qquad \qquad \qquad k\in \mathcal{K}, 1\leq l \leq L_k,
	\end{align}
	where $\theta_{k,l}$ and $\phi_{k,l}$ denote the elevation and azimuth AoDs of the $l$-th channel path between the user $k$ and the BS. Accordingly, the transmit field response vector between $k$-th user and $m$-th FA at BS is given by:
	\begin{equation}
	\mathbf{g}_k(\mathbf{t}_m)=[e^{j\frac{2\pi}{\lambda}\rho(\mathbf{t}_m,\theta_{k,1},\phi_{k,1})},...,e^{j\frac{2\pi}{\lambda}\rho(\mathbf{t}_m,\theta_{k,L_k},\phi_{k,L_k})}]^T,
	\end{equation} 
	where $\lambda$ is the carrier wavelength. Thus, the communication channel between the BS and user $k$ is modeled as follows \cite{b11}:
	\begin{equation}
		\mathbf{h}_k(\tilde{\mathbf{t}})= \mathbf{1}^T_{L_k}\mathbf{\Sigma}_k \mathbf{G}_k(\tilde{\mathbf{t}})\in \mathbb{C}^{N_t\times 1}, \label{channel}
	\end{equation}
	where $\tilde{\mathbf{t}}=[\mathbf{t}_1^T,...,\mathbf{t}_{N_t}^T]^T\in \mathbb{R}^{2N_t \times 1}$ represents the FAs position, $\mathbf{G}_k(\tilde{\mathbf{t}})=[\mathbf{g}_k(\mathbf{t}_1),\mathbf{g}_k(\mathbf{t}_2),...,\mathbf{g}_k(\mathbf{t}_{N_t})]\in \mathbb{C}^{L_k\times N_t}$ represents the field response matrix at the BS, and $\mathbf{\Sigma}_k=\text{diag} \{ [\sigma_{1,k},...,\sigma_{L_k,k}]^T \}$ denotes path response of $L_k$ paths of the channel between the BS and the $k$-th user. 
	
	For the downlink communication, we adopt the per-user signal-to-interference-plus-noise ratio (SINR) to measure the communication QoS. The received signal of $k$-th user is given by:
	\begin{equation}
		y_k = \mathbf{h}_k(\tilde{\mathbf{t}}) \mathbf{x}+z_k, k\in\mathcal{K},
	\end{equation}
	where $\mathbf{h}_k(\tilde{\mathbf{t}})$ is given by (\ref{channel}), and $z_k \sim \mathcal{CN}(0,\sigma_r^2)$ denotes the zero-mean additive white Gaussian noise (AWGN). Then, the SINR of user $k$ is given by:
	\begin{equation}
		\gamma_k(\mathbf{W},\tilde{\mathbf{t}}) = \frac{|\mathbf{h}_k(\tilde{\mathbf{t}}) \mathbf{w}_k|^2}{\sum_{q=1,q\neq k}^{K}|\mathbf{h}_k(\tilde{\mathbf{t}}) \mathbf{w}_q|^2 + \sigma^2}.
	\end{equation}
	
	\subsection{Sensing Model}
	 We adopt the line-of-sight (LoS) channel model for the sensing channel between the BS and the target \cite{N2}. Let $\vartheta$ and $\varphi$ denote the elevation and azimuth angle between the target and the BS, respectively. Since the relatively far distance between the BS and the point target, $\vartheta$ and $\varphi$ are constant at different positions in FAs. Let $\mathbf{a}_r(\vartheta,\varphi)\in \mathbb{C}^{1\times PQ}$ and $\mathbf{a}_t(\vartheta,\varphi,\tilde{\mathbf{t}})\in \mathbb{C}^{1\times N_t}$ denote the receive and transmit steering vectors. Specially, the receive steering vector $\mathbf{a}_r(\vartheta,\varphi)$ is denoted by 
	\begin{equation}
		\mathbf{a}_r(\vartheta,\varphi) = \mathbf{a}_P(\vartheta,\varphi) \otimes \mathbf{a}_Q(\vartheta,\varphi),
	\end{equation}
	where $\mathbf{a}_P(\vartheta,\varphi) = \left[1, e^{j\pi \cos\vartheta \sin\varphi},...,e^{j\pi(P-1)\cos\vartheta \sin\varphi} \right]$, $ \mathbf{a}_Q(\vartheta,\varphi) = \left[1, e^{j\pi \sin\vartheta},...,e^{j\pi(Q-1)\sin\vartheta} \right]$. The transmit steering vector, i.e.,  $\mathbf{a}_t(\theta,\phi,\tilde{\mathbf{t}})$ can be written as
	\begin{equation}
		\mathbf{a}_t(\vartheta,\varphi,\tilde{\mathbf{t}}) = \left[ e^{j\frac{2\pi}{\lambda}\rho(\mathbf{t}_1,\vartheta,\varphi)},...,e^{j\frac{2\pi}{\lambda}\rho(\mathbf{t}_{N_t},\vartheta,\varphi)} \right].
	\end{equation}
	Then, the sensing channel $\mathbf{G}$ can be written as \cite{b5}
	\begin{equation}
		\mathbf{G} \triangleq \alpha \mathbf{a}_r(\vartheta,\varphi)^H 	\mathbf{a}_t(\vartheta,\varphi,\tilde{\mathbf{t}}),  
	\end{equation}
	where $\alpha$ denotes the reflection coefficient of the sensing target. Thus, for radar sensing, the reflected echo signal $\mathbf{y}_r \in \mathbb{C}^{PQ \times 1}$ at the sensing receiver of the BS can be written as :
	\begin{equation}
		\mathbf{y}_r=\mathbf{G}\mathbf{x}+\mathbf{z}_r,
	\end{equation}
	where $\mathbf{z}_r\in \mathbb{C}^{PQ \times 1}\sim\mathcal{CN}(0,\sigma_r^2\mathbf{I}_{PQ})$. Subsequently, the sensing SNR at the sensing receiver which can be written as
	\begin{align}
		& \gamma_s(\mathbf{W},\tilde{\mathbf{t}},\vartheta,\varphi) = \eta \mathbf{a}_t(\vartheta,\varphi,\tilde{\mathbf{t}}) \mathbf{W} \mathbf{W}^H \mathbf{a}_t(\vartheta,\varphi,\tilde{\mathbf{t}})^H, \label{sensing_SNR}
	\end{align}
	where $\eta\triangleq\frac{|\alpha|^2 \mathbf{a}_{r}(\vartheta,\varphi)\mathbf{a}_{r}^H(\vartheta,\varphi)}{\sigma_r^2}$ denotes the integrated coefficient.

	\subsection{Problem Formulation}
	In this paper, we aim to maximize the sensing SNR at the sensing receiver by jointly optimizing the FA position $\tilde{\mathbf{t}}$ and the dual-functional beamforming $\mathbf{W}$ while satisfying the SINR requirements of CUs and the minimum distance between FAs. We consider the following two cases based on the perfect and imperfect CSI of the communication channels and the sensing channels.
	
	\subsubsection{Perfect CSI Case}
	In this case, we assume that the BS has the perfect CSI of the communication channels and the sensing channels. \mcone{ Note that the sensing channel parameters, i.e., $\vartheta$, $\varphi$, and $\alpha$, are known or previously estimated at the BS for designing the dual-functional beamforming and FA positions to detect the point target of interest \cite{b12}.} The corresponding optimization problem is formulated as
	\begin{align}
		\underset{\left\{\mathbf{w}_k\right\}_{k=1}^K,\tilde{\mathbf{t}}}{\text{max}}&~\gamma_s(\mathbf{W},\tilde{\mathbf{t}},\vartheta,\varphi) \label{Problem} \\
		\text{s.t.}&~\gamma_k(\mathbf{W},\tilde{\mathbf{t}}) \geq \Gamma_k,~\forall k, \tag{\ref{Problem}{a}} \label{Problema}\\
		&~\mathbf{t}_m\in \mathcal{C}_t,~1\leq m \leq N_t, \tag{\ref{Problem}{b}} \label{Problemb}\\
		&~\left\| \mathbf{t}_m - \mathbf{t}_n \right\|_2^2 \geq D^2,~1\leq m \neq n \leq N_t, \tag{\ref{Problem}{c}} \label{Problemc}\\
		&~\sum_{k=1}^{K}\text{Tr}(\mathbf{w}_k\mathbf{w}_k^H)\leq P_{max}, \tag{\ref{Problem}{d}} \label{Problemd}
	\end{align}
	where the constraint (\ref{Problema}) denotes the minimum SINR requirement of each user, and the constraint (\ref{Problemb}) indicates the movement region of each FA. The minimum distance $D$ between each pair of FAs in constraint (\ref{Problemc}) is large enough, such that the mutual coupling between adjacent FAs can be reasonably ignored \cite{N1}.
	
	\subsubsection{Imperfect CSI Case}
	In this case, we assume the BS has the imperfect CSI of the communication channels and the sensing channels. For the communication channels, we consider a deterministic CSI error model with channel estimation errors \cite{RM2} \cite{M7}, which is given by
	\begin{equation}
		\mathbf{h}_k(\tilde{\mathbf{t}}) = \widehat{\mathbf{h}}_k(\tilde{\mathbf{t}}) + \boldsymbol{\delta}_k \in \mathbb{C}^{1\times N_t}, \forall k, \label{Imperfect_CSI}
	\end{equation}
	where $\widehat{\mathbf{h}}_k(\tilde{\mathbf{t}})$ is the estimated channel, and $\boldsymbol{\delta}_k \in \mathcal{C}^{1 \times N_t}$ represents the channel estimation error within the region $\mathcal{H}_k = \left\{ \boldsymbol{\delta}_k| \boldsymbol{\delta}_k \mathbf{Q}_i \boldsymbol{\delta}_k^H \leq \varepsilon_k, \mathbf{Q}_i \succeq \mathbf{0}\right\}$. For the sensing channels, we consider a deterministic CSI error model, with the angle estimation errors of the point target, which is given by
	
	\begin{align}
		&\Phi_{\vartheta} \triangleq \left[ \vartheta-\Delta\vartheta, \vartheta+\Delta\vartheta\right],  \nonumber \\
		&\Phi_{\varphi} \triangleq \left[ \varphi-\Delta\varphi, \varphi+\Delta\varphi\right],
	\end{align}
	where $\Delta\varphi$ and $\Delta\varphi$ denote the elevation and azimuth error range, respectively. Accordingly, the corresponding optimization problem is formulated as
	\begin{align}
		\underset{\left\{\mathbf{w}_k\right\}_{k=1}^K,\tilde{\mathbf{t}}}{\text{max}}&~\underset{\vartheta \in \Phi_{\vartheta}, \varphi \in \Phi_{\varphi}}{\text{min}}\gamma_s(\mathbf{W},\tilde{\mathbf{t}},\vartheta,\varphi) \label{Problem2} \\
		\text{s.t.}&~\gamma_k(\mathbf{W},\tilde{\mathbf{t}}) \geq \Gamma_k ,\boldsymbol{\delta}_k \in \mathcal{H}_k , \forall k, \tag{\ref{Problem2}{a}} \label{Problem2a}\\
		&~\mathbf{t}_m\in \mathcal{C}_t,~1\leq m \leq N_t, \tag{\ref{Problem2}{b}} \label{Problem2b}\\
		&~\left\| \mathbf{t}_m - \mathbf{t}_n \right\|_2^2 \geq D^2,~1\leq m \neq n \leq N_t, \tag{\ref{Problem2}{c}} \label{Problem2c}\\
		&~\sum_{k=1}^{K}\text{Tr}(\mathbf{w}_k\mathbf{w}_k^H)\leq P_{max}. \tag{\ref{Problem2}{d}} \label{Problem2d}
	\end{align}
	
	Note that both problem ($\ref{Problem}$) and ($\ref{Problem2}$) are non-convex due to the non-concave objective functions and the non-convex constraints (\ref{Problema}), (\ref{Problemc}), (\ref{Problem2a}), and (\ref{Problem2c}). Besides, the FA position $\tilde{\mathbf{t}}$ and dual-functional beamforming $\mathbf{W}$ are highly coupled in objective functions and constraints, which makes problems ($\ref{Problem}$) and ($\ref{Problem2}$) challenging to solve. In the following two sections, we propose two efficient algorithms to solve these two problems, respectively.
	
	\textit{Remark:} According to the derivation in \cite{M8}, the detector's detection probability can be written as
	\begin{equation}
		P_D = \frac{1}{2}\text{erfc}\left\{ \text{erfc}^{-1}(2P_{FA})-\sqrt{\gamma_r(\mathbf{W},\tilde{\mathbf{t}},\vartheta,\varphi)} \right\},\label{prob_d}
	\end{equation} 
	where $\text{erfc}(x)=\frac{2}{\sqrt{\pi}}\int_{x}^{\infty}e^{-t^2}dt $ denote the error function, and $P_{FA}$ denotes the pre-defined false alarm probability. It can be observed that when $P_{FA}$ is fixed, the target detection probability in ($\ref{prob_d}$) monotonically increases with the sensing SNR. Therefore, maximizing the sensing SNR is equivalent to maximizing the target detection probability.
	
	\section{Proposed design under Perfect CSI}
	In this section, we propose an AO-based algorithm to solve the problem (\ref{Problem}). Specifically, we decompose problem ($\ref{Problem}$) into two sub-problems. For given FA position $\tilde{\mathbf{t}}$, we optimize the dual-functional beamforming $\mathbf{W}$ by SDR. For given dual-functional beamforming $\mathbf{W}$, the FA position $\tilde{\mathbf{t}}$ is optimized by SCA. Finally, we present the overall algorithm and analyze its convergence. 
	
	\subsection{Dual-Functional Beamforming Optimization}
	For any given FA position $\tilde{\mathbf{t}}$, the dual-functional beamforming can be obtained by solving the following problem: 
	\begin{align}
		\underset{\left\{\mathbf{w}_k\right\}_{k=1}^K}{\text{max}}&~\eta \mathbf{a}_t(\vartheta,\varphi,\tilde{\mathbf{t}}) \mathbf{W} \mathbf{W}^H \mathbf{a}_t(\vartheta,\varphi,\tilde{\mathbf{t}})^H \label{Problem3} \\
		\text{s.t.}&~(\ref{Problema}), (\ref{Problemd}) \nonumber.
	\end{align}
	
	Notice that the problem ($\ref{Problem3}$) is non-convex due to the non-convex constraints ($\ref{Problema}$). We then employ SDR technique to solve the problem ($\ref{Problem3}$). Let $\mathbf{T}_k\triangleq\mathbf{w}_k\mathbf{w}_k^H$, $\mathbf{H}_k\triangleq\mathbf{h}_k^H\mathbf{h}_k$, and $\mathbf{A}\triangleq\eta\mathbf{a}_t(\vartheta,\varphi,\tilde{\mathbf{t}})^H\mathbf{a}_t(\vartheta,\varphi,\tilde{\mathbf{t}})$. Problem ($\ref{Problem3}$) can be approximated as
	\begin{align}
		\underset{\left\{\mathbf{T}_k\right\}_{k=1}^K}{\text{max}}&~ \text{Tr}(\mathbf{A}\mathbf{T}_k) \label{Problem4} \\
		\text{s.t.}&\text{Tr}(\mathbf{H}_k\mathbf{T}_k)-\Gamma_k\sum_{q=1,\neq k}^{K}\text{Tr}(\mathbf{H}_k\mathbf{T}_q) \geq \Gamma_k\sigma^2, \forall k,  \nonumber\\
		&~\sum_{k=1}^{K}\text{Tr}(\mathbf{T}_k) \leq P_{max}, \nonumber \\
		&~\mathbf{T}_k \succeq \mathbf{0}, \forall k , \nonumber \\
		&~\text{rank}(\mathbf{T}_k)=1,~\forall k  \tag{\ref{Problem4}{a}} \label{Problem4a}\nonumber.
	\end{align}
	
	Then, by dropping the rank-1 constraints ($\ref{Problem4a}$), the problem (\ref{Problem4}) is relaxed as a standard semidefinite program (SDP) problem, where its optimal solution can be obtained via CVX \cite{b13}. Let $\mathbf{T}_k^*$ denote the optimal solution of the relaxed problem. It is noteworthy that $\mathbf{T}_k^*$ is proved to be rank-1 by \cite[Theorem 2]{b5}. Thus, the optimal solution of the problem (\ref{Problem3}) can be given by:
	
	\begin{equation}
		\mathbf{w}_k^* =\sqrt{\lambda_{max}(\mathbf{T}_k^*)}\mathcal{P}(\mathbf{T}_k^*), \label{decomp}
	\end{equation}
	where $\lambda_{max}(\mathbf{T}_k^*)$ represents the largest eigenvalue of $\mathbf{T}_k^*$, and $\mathcal{P}(\mathbf{T}_k^*)$ is the corresponding eigenvector.
	
	\subsection{FA Position Optimization}
	For any given dual-functional beamforming $\mathbf{W}$, the FA position can be obtained by solving the following problem.
	\begin{align}
		\underset{\tilde{\mathbf{t}}}{\text{max}}&~ \mathbf{a}_t(\vartheta,\varphi,\tilde{\mathbf{t}}) \mathbf{Q} \mathbf{a}_t(\vartheta,\varphi,\tilde{\mathbf{t}})^H \triangleq g(\tilde{\mathbf{t}}) \label{Problem5} \\
		\text{s.t.}&~\frac{|\mathbf{h}_k(\tilde{\mathbf{t}}) \mathbf{w}_k|^2}{\sum_{q=1,q\neq k}^{K}|\mathbf{h}_k(\tilde{\mathbf{t}}) \mathbf{w}_q|^2 + \sigma^2} \geq \Gamma_k,~\forall k, \tag{\ref{Problem5}{a}} \label{Problem5a}\\
		&~\mathbf{t}_m\in \mathcal{C}_t,~1\leq m \leq N_t, \tag{\ref{Problem5}{b}} \label{Problem5b}\\
		&~\left\| \mathbf{t}_m - \mathbf{t}_n \right\|_2^2 \geq D^2,~1\leq m \neq n \leq N_t, \tag{\ref{Problem5}{c}} \label{Problem5c} 
	\end{align}
	where $\mathbf{Q}\triangleq\eta\mathbf{W}\mathbf{W}^H$. Problem (\ref{Problem5}) is non-convex due to the non-concave objective function and the non-convex constraints in (\ref{Problem5a}) and (\ref{Problem5c}), it is challenging to solve the problem (\ref{Problem5}). To deal with the non-convex objective function, (\ref{Problem5a}), and (\ref{Problem5c}), SCA can be applied. 
	
	\subsubsection{Convex Approximation of Objective Function} Let $\tilde{\mathbf{t}}^{\left(r\right)}$ denote the given position of FAs in the $r$-th iteration. Specifically, the objective function is approximated as a more tractable concave function with respect to (w.r.t.) $\tilde{\mathbf{t}}^{\left(r\right)}$ in the $r$-th iteration. According to the second-order Taylor expansion theorem, we approximate the objective function $g(\tilde{\mathbf{t}})$ at $\tilde{\mathbf{t}}^{\left(r\right)}$ \cite{b14} by constructing a concave surrogate function, denoted by:
	\begin{equation}
		g(\tilde{\mathbf{t}})\geq g(\tilde{\mathbf{t}}^{\left(r\right)})+\nabla g(\tilde{\mathbf{t}})^T(\tilde{\mathbf{t}}-\tilde{\mathbf{t}}^{\left(r\right)})-\frac{\delta}{2}(\tilde{\mathbf{t}}-\tilde{\mathbf{t}}^{\left(r\right)})^T(\tilde{\mathbf{t}}-\tilde{\mathbf{t}}^{\left(r\right)}), \label{obj_ste}
	\end{equation}
	where $\nabla g(\tilde{\mathbf{t}}) $ and $\nabla^2 g(\tilde{\mathbf{t}})$ represent the gradient vector and the Hessian matrix of $g(\tilde{\mathbf{t}})$ w.r.t. $\tilde{\mathbf{t}}$, respectively, and $\delta$ is a positive real number satisfying $\delta \mathbf{I}_{2N_t}\succeq \nabla^2 g(\tilde{\mathbf{t}})$. Here, $g(\tilde{\mathbf{t}}^{\left(r\right)})$ is the value of the objective function at $\tilde{\mathbf{t}}^{\left(r\right)}$, $\nabla g(\tilde{\mathbf{t}}^{\left(r\right)})$ ensures the equal gradient between the surrogate function and the original function at $\tilde{\mathbf{t}}^{\left(r\right)}$, and $\delta$ guarantees that the surrogate function is a global lower bound of the original function. The expressions of $\nabla g(\tilde{\mathbf{t}}) $, $\nabla^2 g(\tilde{\mathbf{t}})$, and $\delta$ are shown in Appendix A.
	
	\subsubsection{Convex Approximation of Constraint (\ref{Problem5a})}
	
	The constraint (\ref{Problem5a}) can be transformed into the following form:
	\begin{equation}
		\underbrace{\mathbf{h}_k(\tilde{\mathbf{t}}) \mathbf{R}_k \mathbf{h}_k^H(\tilde{\mathbf{t}})}_{\triangleq f_k(\tilde{\mathbf{t}})} + \Gamma_k \sigma^2 \leq 0, \forall k, \label{f_k}
	\end{equation}
	where $\mathbf{R}_k=\sum_{q=1,q\neq k}^{K}\mathbf{w}_q\mathbf{w}_q^H-\mathbf{w}_k\mathbf{w}_k^H$. As $f_k(\tilde{\mathbf{t}})$ is neither convex nor concave w.r.t. $\tilde{\mathbf{t}}$, we construct a surrogate function that serves as an upper bound of $f_k(\tilde{\mathbf{t}})$ by using the second-order Taylor expansion. Let $\nabla f_k(\tilde{\mathbf{t}}) $ and $\nabla^2 f(\tilde{\mathbf{t}})$ denote the gradient vector and the Hessian matrix of $f_k(\tilde{\mathbf{t}})$ over $\tilde{\mathbf{t}}$. Similarly, we construct a positive real number that satisfies $\zeta_k \mathbf{I}_{2N_t}\succeq \nabla^2 f_k(\tilde{\mathbf{t}})$. Then, we apply second-order Taylor expansion to $f_k(\tilde{\mathbf{t}})$ around $\tilde{\mathbf{t}}^{\left(r\right)}$:
	\begin{equation}
		f_k(\tilde{\mathbf{t}}) \leq f_k(\tilde{\mathbf{t}}^{\left(r\right)}) + \nabla f_k(\tilde{\mathbf{t}}^{\left(r\right)})^T (\tilde{\mathbf{t}} - \tilde{\mathbf{t}}^{\left(r\right)}) +\frac{\zeta_k}{2}(\tilde{\mathbf{t}} - \tilde{\mathbf{t}}^{\left(r\right)})^T(\tilde{\mathbf{t}} - \tilde{\mathbf{t}}^{\left(r\right)}). \label{C4_bound}
	\end{equation} 
	The derivations of $\nabla f(\tilde{\mathbf{t}})$, $\nabla^2 f(\tilde{\mathbf{t}})$, and $\zeta_k$ are shown in Appendix B.
	
	\subsubsection{Convex Approximation of Constraint (\ref{Problem5c})} In constraints (\ref{Problem5c}), since $\left\| \mathbf{t}_m - \mathbf{t}_n \right\|_2^2$ is a convex function w.r.t. $\mathbf{t}_m - \mathbf{t}_n$, we get the following inequality by applying the first-order Taylor expansion at the given point $\mathbf{t}_m^{\left(r\right)}$ and $\mathbf{t}_n^{\left(r\right)}$ \cite{N3}
	\begin{align}
		\left\| \mathbf{t}_m - \mathbf{t}_n \right\|_2^2 \geq -\left\| \mathbf{t}_m^{\left(r\right)} - \mathbf{t}_n^{\left(r\right)} \right\|_2^2 + 2(\mathbf{t}_m^{\left(r\right)} - \mathbf{t}_n^{\left(r\right)})^T  \nonumber \\
		\times (\mathbf{t}_m - \mathbf{t}_n),~1\leq m \neq n \leq N_t. \label{C3_bound}
	\end{align}
	
	Thus, at the $r$-th iteration, by using the lower bounds in ($\ref{C3_bound}$) and upper bounds in ($\ref{C4_bound}$), the problem ($\ref{Problem5}$) is approximated as the following problem:
	\begin{align}
		\underset{\tilde{\mathbf{t}}}{\text{max}}&~g(\tilde{\mathbf{t}}^{\left(r\right)})+\nabla g(\tilde{\mathbf{t}})^T(\tilde{\mathbf{t}}-\tilde{\mathbf{t}}^{\left(r\right)})-\frac{\delta}{2}(\tilde{\mathbf{t}}-\tilde{\mathbf{t}}^{\left(r\right)})^T(\tilde{\mathbf{t}}-\tilde{\mathbf{t}}^{\left(r\right)}) \label{Problem6} \\
		\text{s.t.}
		&~\mathbf{t}_m\in \mathcal{C}_t,~1\leq m \leq N_t, \tag{\ref{Problem6}{a}} \label{Problem6a}\\		
		&~D^2 \leq -\left\| \mathbf{t}_m^{\left(r\right)} - \mathbf{t}_n^{\left(r\right)} \right\|_2^2 + 2(\mathbf{t}_m^{\left(r\right)} - \mathbf{t}_n^{\left(r\right)})^T  \nonumber \\
		&~\times(\mathbf{t}_m - \mathbf{t}_n),~1\leq m \neq n \leq N_t, \tag{\ref{Problem6}{b}} \label{Problem6b}\\
		&~f_k(\tilde{\mathbf{t}}^{\left(r\right)}) + \nabla f_k(\tilde{\mathbf{t}}^{\left(r\right)})^T (\tilde{\mathbf{t}} - \tilde{\mathbf{t}}^{\left(r\right)}) \nonumber \\
		&~+\frac{\zeta_k}{2}(\tilde{\mathbf{t}} - \tilde{\mathbf{t}}^{\left(r\right)})^T(\tilde{\mathbf{t}} - \tilde{\mathbf{t}}^{\left(r\right)}) + \Gamma_k \sigma^2 \leq 0, \forall k.\tag{\ref{Problem6}{c}} \label{Problem6c}
	\end{align}
	
	It can be observed that the objective function and constraints of the problem ($\ref{Problem6}$) are convex. Therefore, the problem ($\ref{Problem6}$) is a convex optimization problem, which can be efficiently solved with existing optimization tools \cite{b13}.
	
	\subsection{Overall Algorithm and Convergence}
	Once obtaining the solutions of the problems ($\ref{Problem4}$) and ($\ref{Problem6}$), we can proceed with the completion of our proposed AO-based iterate algorithm for solving the problem ($\ref{Problem}$). Let $\mathbf{W}^{\left(r\right)}$ denote the optimal solution of the problem (\ref{Problem3}) in the $r$-th iteration. In the $r$-th iteration, we obtain $\mathbf{W}^{\left(r+1\right)}$ by solving the problem ($\ref{Problem4}$) while keeping $\tilde{\mathbf{t}}^r$ fixed. Then, for given $\mathbf{W}^{\left(r+1\right)}$, we obtain $\tilde{\mathbf{t}}^{\left(r+1\right)}$ by solving the problem ($\ref{Problem6}$). The algorithm iteratively solves the two sub-problems until the increase of the sensing SNR in ($\ref{sensing_SNR}$) falls below a predefined convergence threshold $\xi$. The overall algorithm is summarized in Algorithm~1.
	
	\begin{algorithm}[t!]
		\caption{Proposed Algorithm for Solving Problem ($\ref{Problem}$)}
		\begin{algorithmic}[1]
			\renewcommand{\algorithmicrequire}{\textbf{Input:}}
			\renewcommand{\algorithmicensure}{\textbf{Output:}}
			\REQUIRE $N_t$, $PQ$, $K$, $\mathcal{C}_t$, $\{L_k\}_{k=1}^K$, $\{ \Sigma_k \}_{k=1}^K$, $\sigma^2$, $\{\theta_{k,l}\}$, $\{\phi_{k,l}\}$, $\theta$, $\phi$, $\{\Gamma_k\}$, $P_{max}$, $D$, $\xi$.
			\STATE Let $r=0$. Initialize the FA positions $\tilde{\mathbf{t}}^{\left(0\right)}$. 
			\REPEAT
			\STATE Given $\tilde{\mathbf{t}}^{\left(r\right)}$, obtain $\mathbf{W}^{\left(r+1\right)}$ by solving the problem ($\ref{Problem3}$) with the SDP technique.
			\STATE Given $\mathbf{W}^{\left(r+1\right)}$, obtain $\tilde{\mathbf{t}}^{\left(r+1\right)}$ by solving the problem ($\ref{Problem5}$) with the SCA technique.
			\STATE Update $r = r + 1$
			\UNTIL{$\text{the value of the sensing SNR in}~(\ref{sensing_SNR})~\text{converged}~\text{or}$ \\ $\text{the maximum number of iterations reached}$}
		\end{algorithmic}
	\end{algorithm}
	
	\textit{Proposition 1:} The value of the objective function in the problem ($\ref{Problem}$) is non-decreasing with iterations, and Algorithm~1 converges.
	
	\textit{Proof:}~Please refer to Appendix C.

	\section{Proposed design under Imperfect CSI}
	Compared to the problem ($\ref{Problem}$), the problem ($\ref{Problem2}$) is more complicated due to the max-min optimization and infinite number of inequalities involved in (\ref{Problem2a}). Thus, the previous algorithm is no longer applicable. In this section, we propose an AO-based algorithm to solve the problem ($\ref{Problem2}$). To begin with, \mcone{let $\mathbf{T}_k\triangleq \mathbf{w}_k \mathbf{w}_k^H$}, and we introduce auxiliary variable $z$ to transform the max-min problem ($\ref{Problem2}$) into the following equivalent form:
	\begin{align}
		\underset{\left\{\mathbf{T}_k\right\},\tilde{\mathbf{t}},z}{\text{max}}&~z \label{Problem7} \\
		\text{s.t.}&~ \gamma_s(\left\{\mathbf{T}_k\right\},\tilde{\mathbf{t}},\vartheta,\varphi) \geq z, \vartheta \in \Phi_{\vartheta}, \varphi \in \Phi_{\varphi}, \tag{\ref{Problem7}{a}} \label{Problem7a} \\
		&~(\ref{Problem2a}),(\ref{Problem2b}), (\ref{Problem2c}), (\ref{Problem2d}).\nonumber
	\end{align}
	
	Then, the problem (\ref{Problem7}) is decomposed into two sub-problems: for given FA position $\tilde{\mathbf{t}}$, we optimize the dual-functional beamforming $\left\{\mathbf{T}_k\right\}$ by $\mathcal{S}$-procedure and SDP; and for given $\left\{\mathbf{T}_k\right\}$, we optimize $\tilde{\mathbf{t}}$ by SCA and $\mathcal{S}$-procedure. Finally, the overall algorithm is presented. 
	
	\subsection{Dual-Functional Beamforming Optimization}
	For any given FA position $\tilde{\mathbf{t}}$, the subproblem is given by
	\begin{align}
		\underset{\left\{\mathbf{T}_k\right\},z}{\text{max}}&~z \label{Problem8} \\
		\text{s.t.}&~ \gamma_s(\left\{\mathbf{T}_k\right\},\tilde{\mathbf{t}},\vartheta,\varphi) \geq z, \vartheta \in \Phi_{\vartheta}, \varphi \in \Phi_{\varphi}, \tag{\ref{Problem8}{a}} \label{Problem8a} \\
		&~(\ref{Problem2a}),(\ref{Problem2d}).\nonumber
	\end{align}
	
	Likewise, recall problem ($\ref{Problem3}$), the SDP technique is adopted. Let $\mathbf{A}(\vartheta,\varphi)\triangleq\eta\mathbf{a}_t(\vartheta,\varphi,\tilde{\mathbf{t}})^H\mathbf{a}_t(\vartheta,\varphi,\tilde{\mathbf{t}})$. Substituting (\ref{Imperfect_CSI}) into ($\ref{Problem2a}$), the problem ($\ref{Problem8}$) can be rewritten as
	\begin{align}
		\underset{\mathbf{T}_k,z}{\text{max}}&~z \label{Problem9} \\
		\text{s.t.}&~ \sum_{k=1}^{K}\text{Tr}\left(\mathbf{A}(\vartheta,\varphi)\mathbf{T}_k\right) \geq z, \vartheta \in \Phi_{\vartheta}, \varphi \in \Phi_{\varphi}, \tag{\ref{Problem9}{a}} \label{Problem9a} \\
		&~ -\boldsymbol{\delta}_k\mathbf{R}_k\boldsymbol{\delta}_k^H -2\text{Re}\left\{\boldsymbol{\delta}_k\mathbf{R}_k\widehat{\mathbf{h}}_k(\tilde{\mathbf{t}})^H\right\} \nonumber - \widehat{\mathbf{h}}_k(\tilde{\mathbf{t}})\mathbf{R}_k\widehat{\mathbf{h}}_k(\tilde{\mathbf{t}})^H \nonumber\\
		&\qquad\qquad\qquad+ \gamma_k \sigma^2 \leq 0, \boldsymbol{\delta}_k \in \mathcal{H}_k , \forall k, \tag{\ref{Problem9}{b}} \label{Problem9b} \\
		&~\sum_{k=1}^{K}\text{Tr}(\mathbf{T}_k) \leq P_{max}, \tag{\ref{Problem9}{c}} \label{Problem9c} \\
		&~\mathbf{T}_k \succeq \mathbf{0}, \forall k , \tag{\ref{Problem9}{d}} \label{Problem9d} \\
		&~\text{rank}(\mathbf{T}_k)=1,~\forall k  \tag{\ref{Problem9}{e}} \label{Problem9e},
	\end{align}
	where $\mathbf{R}_k=\sum_{q=1,q\neq k}^{K}\mathbf{T}_q-\mathbf{T}_k$. To tickle the number of inequality constraints in ($\ref{Problem9b}$), we apply the following lemma to convert ($\ref{Problem9b}$) into easier cases.
	
	\textit{Lemma 1 ($\mathcal{S-}Procedure$\cite{M10}):} Let $f_i(\mathbf{x})=\mathbf{x}^H\mathbf{A}_i\mathbf{x}+2\text{Re}\left\{\mathbf{b}_i\mathbf{x}\right\}+c_i$, $i\in\left\{1,2\right\}$, where $\mathbf{A}_i\in\mathbb{C}^{N\times N}$ is a complex Hermitian matrices,  $\mathbf{x}\in \mathbb{C}^{N\times 1}$, and $\mathbf{b}_i\in \mathbb{C}^{N\times 1}$. The condition $f_1(\mathbf{x})\geq 0 \Rightarrow f_2(\mathbf{x})\geq 0$ holds if and only if there exists $\nu\geq 0$ such that
	\begin{equation}
		\begin{bmatrix}
			\mathbf{A}_2 & \mathbf{b}_2 \\ \mathbf{b}_2^H & c_2
		\end{bmatrix}
		-
		\nu \begin{bmatrix}
			\mathbf{A}_1 & \mathbf{b}_1 \\ \mathbf{b}_1^H & c_1
		\end{bmatrix}
		\succeq \mathbf{0}.
	\end{equation}
	By applying Lemma 1, the implication $\boldsymbol{\delta}_k\mathbf{Q}_k\boldsymbol{\delta}_k^H - \varepsilon_k \leq 0 \Rightarrow (\ref{Problem9b})$ holds if and only if there exists a nonnegative $\nu_k$ such that 
	\begin{align}
		&\nu_k\begin{bmatrix}
			\mathbf{Q}_k & \mathbf{0} \\ \mathbf{0} & -\varepsilon_k
		\end{bmatrix}
		-
		\begin{bmatrix}
			-\mathbf{R}_k & -\mathbf{R}_k\widehat{\mathbf{h}}_k(\tilde{\mathbf{t}})^H \\ -\widehat{\mathbf{h}}_k(\tilde{\mathbf{t}})\mathbf{R}_k^H & -\widehat{\mathbf{h}}_k(\tilde{\mathbf{t}})\mathbf{R}_k\widehat{\mathbf{h}}_k(\tilde{\mathbf{t}})^H+\gamma_k \sigma^2
		\end{bmatrix} \nonumber\\
		&\qquad\qquad\qquad\qquad\qquad\qquad\qquad\qquad\succeq \mathbf{0}_{N_t+1}.
	\end{align}
	
	Hence, the problem ($\ref{Problem9}$) can be relaxed into a SDP problem by discarding the rand-1 constraints ($\ref{Problem9e}$), which is given by:
	\begin{align}
		\underset{\left\{\mathbf{T}_k\right\},z,\{\nu_k\}}{\text{max}}&~z \label{Problem10} \\
		\text{s.t.}&~ (\ref{Problem9a}),(\ref{Problem9c}),(\ref{Problem9d}),(\ref{Problem9e}), \nonumber\\
		&\begin{bmatrix}
			\nu_k\mathbf{Q}_k+\mathbf{R}_k & \mathbf{R}_k\widehat{\mathbf{h}}_k(\tilde{\mathbf{t}})^H \\ \widehat{\mathbf{h}}_k(\tilde{\mathbf{t}})\mathbf{R}_k^H & \widehat{\mathbf{h}}_k(\tilde{\mathbf{t}})\mathbf{R}_k\widehat{\mathbf{h}}_k(\tilde{\mathbf{t}})^H-\gamma_k \sigma^2-\varepsilon_k
		\end{bmatrix} \nonumber \\
		&\qquad\qquad\qquad\qquad\qquad\qquad\succeq \mathbf{0}_{N_t+1},\forall k\tag{\ref{Problem10}{a}} \label{Problem10a},		\\
		& \nu_k \geq 0, \forall k. \tag{\ref{Problem10}{b}}
	\end{align}
	
	The above SDP problem can be efficiently solved by standard convex optimization solvers \cite{b13}. 
	
	\subsection{FA Position Optimization}
	Let $\mathbf{M}\triangleq\sum_{k=1}^{K}\mathbf{T}_k$. For any given \mcone{$\left\{\mathbf{T}_k\right\}$}, the subproblem can be written as
	\begin{align}
		\underset{\tilde{\mathbf{t}},z}{\text{max}}&~z \label{Problem11} \\
		\text{s.t.}&~\underbrace{\mathbf{a}_t(\vartheta,\varphi,\tilde{\mathbf{t}})\mathbf{M}\mathbf{a}_t(\vartheta,\varphi,\tilde{\mathbf{t}})^H}_{\triangleq u(\tilde{\mathbf{t}},\vartheta,\varphi)}  \geq z, \vartheta \in \Phi_{\vartheta}, \varphi \in \Phi_{\varphi}, \tag{\ref{Problem11}{a}} \label{Problem11a} \\
		&~(\ref{Problem2a}),(\ref{Problem2b}),(\ref{Problem2c}).\nonumber
	\end{align}
	Note that the problem ($\ref{Problem11}$) is nonconvex due to the nonconvex constraints. To tackle the nonconvexity of ($\ref{Problem11a}$),($\ref{Problem2a}$), and ($\ref{Problem2c}$), we apply SCA where in each iteration, the non-convex sets are replaced by approximating convex constraints at a local point. 
	
	\subsubsection{Convex Approximation of Constraint (\ref{Problem11a})} In the $j$-th iteration, let $\tilde{\mathbf{t}}^{\left(j\right)}$ denote the given position of FAs. To handle the non-convex constraint ($\ref{Problem11a}$), we take the second-order Taylor expansion of $u(\tilde{\mathbf{t}},\vartheta,\varphi)$ at $\tilde{\mathbf{t}}^{\left(j\right)}$ to obtain the following lower bound:
	\begin{align}
		&u(\tilde{\mathbf{t}},\vartheta,\varphi)\geq u(\tilde{\mathbf{t}}^{\left(j\right)},\vartheta,\varphi) + \nabla u(\tilde{\mathbf{t}}^{\left(j\right)},\vartheta,\varphi)\left(\tilde{\mathbf{t}}-\tilde{\mathbf{t}}^{\left(j\right)}\right)^T\nonumber\\
		&\qquad\qquad\qquad\qquad-\frac{\delta_2}{2}\Vert \tilde{\mathbf{t}}-\tilde{\mathbf{t}}^{\left(j\right)} \label{p11a_l} \Vert_2^2,
	\end{align}
	where $\nabla u(\tilde{\mathbf{t}},\vartheta,\varphi)$ represents the gradient vector of $u(\tilde{\mathbf{t}},\vartheta,\varphi)$ w.r.t. $\tilde{\mathbf{t}}$, $\delta_2$ guarantees the approximate function to be a global lower bound of $u(\tilde{\mathbf{t}},\vartheta,\varphi)$. Note that the derivations of $\nabla u(\tilde{\mathbf{t}},\vartheta,\varphi)$ and $\delta_2$ are similar to the derivations of $\nabla g(\tilde{\mathbf{t}})$ in (\ref{delta_g}) and $\delta$ in (\ref{delta}), which are given in Appendix~A.
	
	\subsubsection{Convex Approximation of Constraint (\ref{Problem2a})} Substituting $\mathbf{h}_k(\tilde{\mathbf{t}}) = \widehat{\mathbf{h}}_k(\tilde{\mathbf{t}}) + \boldsymbol{\delta}_k$ into ($\ref{Problem2a}$), we can equivalently rewrite ($\ref{Problem2a}$) as:
	\begin{align}
		&~ \boldsymbol{\delta}_k\mathbf{R}_k\boldsymbol{\delta}_k^H +\underbrace{2\text{Re}\left\{\boldsymbol{\delta}_k\mathbf{R}_k\widehat{\mathbf{h}}_k(\tilde{\mathbf{t}})^H\right\}}_{\triangleq w_k(\tilde{\mathbf{t}})}+ \underbrace{\widehat{\mathbf{h}}_k(\tilde{\mathbf{t}})\mathbf{R}_k\widehat{\mathbf{h}}_k(\tilde{\mathbf{t}})^H}_{\triangleq d_k(\tilde{\mathbf{t}})} \nonumber\\
		&\qquad\qquad\qquad- \gamma_k \sigma^2 \geq 0, \boldsymbol{\delta}_k \in \mathcal{H}_k ,\label{app}
	\end{align}
	where $w_k(\tilde{\mathbf{t}}) \triangleq \boldsymbol{\delta}_k\mathbf{R}_k\widehat{\mathbf{h}}_k(\tilde{\mathbf{t}})^H$, $d_k(\tilde{\mathbf{t}}) \triangleq \widehat{\mathbf{h}}_k(\tilde{\mathbf{t}})\mathbf{R}_k\widehat{\mathbf{h}}_k(\tilde{\mathbf{t}})^H$. Note that $2\text{Re}\left\{\boldsymbol{\delta}_k\mathbf{R}_k\widehat{\mathbf{h}}_k(\tilde{\mathbf{t}})^H\right\}$ and $\widehat{\mathbf{h}}_k(\tilde{\mathbf{t}})\mathbf{R}_k\widehat{\mathbf{h}}_k(\tilde{\mathbf{t}})^H$ are neither convex nor concave w.r.t. $\tilde{\mathbf{t}}$, thus the constraint ($\ref{Problem2a}$) is not a convex set. To tackle the non-convexity of $\widehat{\mathbf{h}}_k(\tilde{\mathbf{t}})\mathbf{R}_k\widehat{\mathbf{h}}_k(\tilde{\mathbf{t}})^H$, we approximate the original function with a concave function w.r.t. $\tilde{\mathbf{t}}$, which is given by
	\begin{align}
		&~d_k(\tilde{\mathbf{t}}) \leq d_k(\tilde{\mathbf{t}}^{\left(j\right)}) + \nabla d_k(\tilde{\mathbf{t}}^{\left(j\right)})^T (\tilde{\mathbf{t}} - \tilde{\mathbf{t}}^{\left(j\right)})\nonumber \\
	 	&~\qquad \qquad \qquad-\frac{\omega_k}{2}(\tilde{\mathbf{t}} - \tilde{\mathbf{t}}^{\left(j\right)})^T(\tilde{\mathbf{t}} - \tilde{\mathbf{t}}^{\left(j\right)}). \label{p11_2}
	\end{align}  
	The derivations of $\nabla d_k(\tilde{\mathbf{t}})$ and $\omega_k$ are similar to $\nabla f_k(\tilde{\mathbf{t}})$ and $\zeta_k$ in (\ref{delte_f_k}) and (\ref{delta}), referring to Appendix~B.
	
	Then, for the nonconvex term $2\text{Re}\left\{\boldsymbol{\delta}_k\mathbf{R}_k\widehat{\mathbf{h}}_k(\tilde{\mathbf{t}})^H\right\}$, we take the second-order Taylor expansion to get the following lower bound:
	\begin{equation}
		w_k(\tilde{\mathbf{t}})\geq w_k(\tilde{\mathbf{t}}^{\left(j\right)}) + \nabla w_k(\tilde{\mathbf{t}}^{\left(j\right)})\left(\tilde{\mathbf{t}}-\tilde{\mathbf{t}}^{\left(j\right)}\right)^T-\frac{\chi_k}{2}\Vert \tilde{\mathbf{t}}-\tilde{\mathbf{t}}^{\left(j\right)} \Vert_2^2, \label{app_2}
	\end{equation}
	where $\nabla w_k(\tilde{\mathbf{t}})$ denotes the gradient vector of $w_k(\tilde{\mathbf{t}})$, $\chi_k$ is a positive real number satisfying $\chi_k \mathbf{I}_{2N_t}\succeq \nabla^2 w_k(\tilde{\mathbf{t}})$. The derivations of $\nabla w_k(\tilde{\mathbf{t}})$ and $\nabla^2 w_k(\tilde{\mathbf{t}})$ are given in Appendix D. Then, substituting ($\ref{C4_bound}$) and ($\ref{app_2}$) into ($\ref{app}$), we have the following approximating convex constraint:
 	\begin{align}
 		&~ \boldsymbol{\delta}_k\mathbf{R}_k\boldsymbol{\delta}_k^H +w_k(\tilde{\mathbf{t}}^{\left(j\right)}) + \nabla w_k(\tilde{\mathbf{t}}^{\left(j\right)})\left(\tilde{\mathbf{t}}-\tilde{\mathbf{t}}^{\left(j\right)}\right)^T \nonumber \\
 		&~-\frac{\chi_k}{2}\Vert \tilde{\mathbf{t}}-\tilde{\mathbf{t}}^{\left(j\right)} \Vert_2^2 +\widehat{\mathbf{h}}_k(\tilde{\mathbf{t}}^{\left(j\right)})\mathbf{R}_k\widehat{\mathbf{h}}_k(\tilde{\mathbf{t}}^{\left(j\right)})^H \nonumber \\ 
 		& + \nabla f(\tilde{\mathbf{t}}^{\left(j\right)})\left(\tilde{\mathbf{t}}-\tilde{\mathbf{t}}^{\left(j\right)}\right)^T-\frac{\zeta_k}{2}\Vert \tilde{\mathbf{t}}-\tilde{\mathbf{t}}^{\left(j\right)} \Vert_2^2  \nonumber\\
 		&\qquad\qquad\qquad- \gamma_k \sigma^2 \geq 0, \boldsymbol{\delta}_k \in \mathcal{H}_k .\label{app_in}
 	\end{align}
 	
 	Then, to facilitate the algorithm design, we rewrite $\nabla w(\tilde{\mathbf{t}}^{\left(r\right)})$ in ($\ref{par_w}$) into a compact form below:
 	\begin{equation}
 		\nabla w_k(\tilde{\mathbf{t}}^{\left(j\right)})=\boldsymbol{\delta}_k\mathbf{z}_k(\tilde{\mathbf{t}},\tilde{\mathbf{t}}^{\left(j\right)})+\mathbf{z}_k(\tilde{\mathbf{t}},\tilde{\mathbf{t}}^{\left(j\right)})^H\boldsymbol{\delta}_k^H,
 	\end{equation}
 	where
 	\begin{align}
 		&~\mathbf{z}_k(\tilde{\mathbf{t}},\tilde{\mathbf{t}}^{\left(j\right)})\triangleq \nonumber \\ &~\left[\mathbf{s}(\mathbf{t}_1^{\left(j\right)}),\mathbf{p}(\mathbf{t}_1^{\left(j\right)}),...,\mathbf{s}(\mathbf{t}_{N_t}^{\left(j\right)}),\mathbf{p}(\mathbf{t}_{N_t}^{\left(j\right)})\right]\left(\tilde{\mathbf{t}}-\tilde{\mathbf{t}}^{\left(j\right)}\right)^T.
 	\end{align}	
 	
 	Thus, ($\ref{app_in}$) can be written into a more traceable form, which is given by
 	\begin{align}
 			\begin{cases}
 				 &\boldsymbol{\delta}_k\mathbf{R}_k\boldsymbol{\delta}_k^H +2\text{Re}\left\{\boldsymbol{\delta}_k\left(\mathbf{R}_k\widehat{\mathbf{h}}_k(\tilde{\mathbf{t}}^{\left(j\right)})^H+\mathbf{z}_k(\tilde{\mathbf{t}},\tilde{\mathbf{t}}^{\left(j\right)})\right)\right\}\\
 				 &\qquad\qquad\qquad\qquad\qquad\qquad\qquad +c_k(\tilde{\mathbf{t}})-\gamma_k \sigma^2 \geq 0,\\
 				&\boldsymbol{\delta}_k\mathbf{Q}_k\boldsymbol{\delta}_k^H - \varepsilon_k \leq 0,\label{rob}
 			\end{cases}
 	\end{align}
 	where $c_k(\tilde{\mathbf{t}})\triangleq \widehat{\mathbf{h}}_k(\tilde{\mathbf{t}}^{\left(j\right)})\mathbf{R}_k\widehat{\mathbf{h}}_k(\tilde{\mathbf{t}}^{\left(j\right)})^H + \nabla f(\tilde{\mathbf{t}}^{\left(j\right)})\left(\tilde{\mathbf{t}}-\tilde{\mathbf{t}}^{\left(j\right)}\right)^T-\frac{(\zeta_k+\chi_k)}{2}\Vert \tilde{\mathbf{t}}-\tilde{\mathbf{t}}^{\left(j\right)} \Vert_2^2$. Thus, based on Lemma 1, constraint ($\ref{rob}$) is equivalently expressed as
 	\begin{align}
 		&\begin{bmatrix}
 			\lambda_k\mathbf{Q}_k+\mathbf{R}_k & \mathbf{R}_k\widehat{\mathbf{h}}_k(\tilde{\mathbf{t}^{\left(j\right)}})^H+\mathbf{z}_k(\tilde{\mathbf{t}},\tilde{\mathbf{t}}^{\left(j\right)}) \\ \widehat{\mathbf{h}}_k(\tilde{\mathbf{t}})\mathbf{R}_k^H + \mathbf{z}_k(\tilde{\mathbf{t}},\tilde{\mathbf{t}}^{\left(j\right)})^H& c_k(\tilde{\mathbf{t}})-\gamma_k \sigma^2-\varepsilon_k
 		\end{bmatrix} \nonumber \\
 		&\qquad\qquad\qquad\qquad\qquad\qquad\succeq \mathbf{0}_{N_t+1}\label{vex}
 	\end{align}
 	where $\lambda_k$ is a nonnegative variable. However, constraint ($\ref{vex}$) is nonconvex due to the existence of the second-order variable $\Vert \tilde{\mathbf{t}}-\tilde{\mathbf{t}}^{\left(j\right)} \Vert_2^2$. To overcome this difficulty, the following lemma can be used.
 	
	\textit{Lemma 2 \cite{M11}:} For any complex Hermitian matrix $\mathbf{A}\in\mathbb{C}^{N\times N}$, $\mathbf{b}\in \mathbb{C}^{N\times 1}$, $c\in \mathbb{R}$, and $d\in \mathbb{R}$, \mcone{it follows that}
	\begin{equation}
		\begin{bmatrix}
			\mathbf{A} & \mathbf{b} \\ \mathbf{b}^H & c
		\end{bmatrix}
		\succeq \mathbf{0}
		\Rightarrow
		\begin{bmatrix}
			\mathbf{A} & \mathbf{b} \\ \mathbf{b}^H & d
		\end{bmatrix}
		\succeq \mathbf{0},
	\end{equation}
	holds if and only if $d\geq c$.

	By applying the Lemma 2, constraint ($\ref{vex}$) can be expressed as the following convex constraints
	\begin{align}
	&\begin{bmatrix}
		\lambda_k\mathbf{Q}_k+\mathbf{R}_k & \mathbf{R}_k\widehat{\mathbf{h}}_k(\tilde{\mathbf{t}^{\left(j\right)}})^H+\mathbf{z}_k(\tilde{\mathbf{t}},\tilde{\mathbf{t}}^{\left(j\right)}) \\ \widehat{\mathbf{h}}_k(\tilde{\mathbf{t}})\mathbf{R}_k^H + \mathbf{z}_k(\tilde{\mathbf{t}},\tilde{\mathbf{t}}^{\left(j\right)})^H& \beta_k-\gamma_k \sigma^2-\varepsilon_k
	\end{bmatrix} \nonumber \\
	&\qquad\qquad\qquad\qquad\qquad\qquad\succeq \mathbf{0}_{N_t+1},\label{vex_convex}
	\end{align}
	where $\beta_k$ is an auxiliary variable satisfying $\beta_k\leq c_k(\tilde{\mathbf{t}})$.
	
	\subsubsection{Convex Approximation of ($\ref{Problem2c}$)} For the nonconvex constraints ($\ref{Problem2c}$), the approximating convex form ($\ref{C3_bound}$) can be applied. 
	
	Then, at the $j$-th iteration, by substituting (\ref{p11a_l}), (\ref{vex_convex}), (\ref{Problem12b}) and (\ref{C3_bound}) into the problem (\ref{Problem11}), we have
	\begin{align}
		\underset{\tilde{\mathbf{t}},z,{\lambda_k},{\beta_k}}{\text{max}}&~z \label{Problem12} \\
		\text{s.t.}&~ u(\tilde{\mathbf{t}}^{\left(j\right)},\vartheta,\varphi) + \nabla u(\tilde{\mathbf{t}}^{\left(j\right)},\vartheta,\varphi)\left(\tilde{\mathbf{t}}-\tilde{\mathbf{t}}^{\left(j\right)}\right)^T\nonumber\\
		&\qquad\qquad\qquad-\frac{\delta}{2}\Vert \tilde{\mathbf{t}}-\tilde{\mathbf{t}}^{\left(j\right)} \Vert_2^2 \geq z, \vartheta \in \Phi_{\vartheta}, \varphi \in \Phi_{\varphi}, \tag{\ref{Problem12}{a}} \label{Problem12a} \\
		& \beta_k \leq c_k(\tilde{\mathbf{t}}),\forall k,\tag{\ref{Problem12}{b}} \label{Problem12b}\\
		& \lambda_k \geq 0,\forall k,\tag{\ref{Problem12}{c}} \label{Problem12c}\\
		&(\ref{vex_convex}), (\ref{C3_bound}),(\ref{Problem2b}).\nonumber
	\end{align}
	
	Since the constraint function in (\ref{Problem12b}) is concave w.r.t. $\tilde{\mathbf{t}}$, implying (\ref{Problem12b}) is a convex set, the problem (\ref{Problem12}) is a convex optimization problem. Thus, it can be efficiently solved by using the standard convex optimization method \cite{b13}. 
	
	\subsection{Overall Algorithm and Convergence}
	This AO-based Algorithm is organized as follows: Firstly, ($\tilde{\mathbf{t}}^{\left(0\right)}$, $\left\{\mathbf{T}_k\right\}^{\left(0\right)}$) are initialized at the beginning of the iteration. Secondly, ($\tilde{\mathbf{t}}^{\left(j\right)}$, $\left\{\mathbf{T}_k\right\}^{\left(j\right)}$) are updated with iteration numbers. Then, the second step is conducted repeatedly until Algorithm 2 converges. \mcone{Finally, we obtain the converged solution of the problem ($\ref{Problem7}$). The details of the proposed algorithm for solving the problem ($\ref{Problem7}$) are summarized in Algorithm 2.}
	
	\textit{Proposition 2:} The value of the objective function in the problem ($\ref{Problem7}$) is non-decreasing with iterations, and Algorithm~2 converges.
	
	\textit{Proof:}~Please refer to Appendix E.
	
	Furthermore, once obtaining a converged solution of the  problem~(\ref{Problem7}), we can construct a solution of the problem (\ref{Problem2}) by the converged solution of the problem~(\ref{Problem7}) with the Gaussian randomization method. Note that the constructed solution for the problem (\ref{Problem2}) is near-optimal \cite{M12}. 

	\begin{algorithm}[t!]
		\caption{Proposed Algorithm for Solving Problem ($\ref{Problem2}$)}
		\begin{algorithmic}[1]
			\renewcommand{\algorithmicrequire}{\textbf{Input:}}
			\renewcommand{\algorithmicensure}{\textbf{Output:}}
			\REQUIRE $N_t$, $PQ$, $K$, $\mathcal{C}_t$, $\{L_k\}_{k=1}^K$, $\{ \Sigma_k \}_{k=1}^K$, $\sigma^2$, $\{\theta_{k,l}\}$, $\{\phi_{k,l}\}$, $\theta$, $\phi$, $\{\Gamma_k\}$, $P_{max}$, $D$, $\xi$,$\{\varepsilon_k\}$,$\Phi_{\vartheta}$,$\Phi_{\varphi}$.
			\STATE Initialize $\tilde{\mathbf{t}}^{\left(0\right)}$ and $\left\{\mathbf{T}_k\right\}^{\left(0\right)}$. Let $j=0$.
			\REPEAT
			\STATE Given $\tilde{\mathbf{t}}^{\left(j\right)}$, solve the problem ($\ref{Problem8}$) by using Lemma 1 and SDP techniques to obtain $\left\{\mathbf{T}_k\right\}^{\left(j+1\right)}$.
			\STATE Given $\left\{\mathbf{T}_k\right\}^{\left(j+1\right)}$, solve the problem ($\ref{Problem11}$) by using SCA, Lemma 1, and Lemma 2 to obtain $\tilde{\mathbf{t}}^{\left(j+1\right)}$.
			\STATE Update $j = j + 1$.
			\UNTIL{$\text{Convergence}$} 
		\end{algorithmic}
	\end{algorithm}
	\section{Simulation Results}
	In this section, we numerically evaluate the performances of the proposed designs under perfect CSI, namely Prop.P-FA, and imperfect CSI, namely Prop.IP-FA. We set $N_t=4$, $P=Q=2$, $K=4$, $\lambda=0.06$ meter (m), $\vartheta = 45^{\circ}$, $\varphi=-30^{\circ}$, and $\xi = 10^{-3}$ if not specified otherwise. The wave The moving region for FAs is set as a square area of size $A\times A$. We set minimum SINR requirements of each CU $\gamma_1=\cdots=\gamma_K=\gamma=10dB$. The users are randomly distributed around the BS, and the distances follow uniform distributions, i.e., $d_k\sim \mathcal{U}[20,100]$, $k\in \mathcal{K}$. We set the numbers of receive paths for each user to be the same, i.e., $L_k=12, k\in \mathcal{K}$. All path responses are i.i.d. circularly symmetric complex Gaussian (CSCG) random variables, i.e., $\sigma_{l,k}\sim \mathcal{CN}(0,\rho d_k^{-\alpha_0}/L)$, where $\rho d_k^{-\alpha_0}$ is the expected channel power gain of user $k$, $\rho=-40$ dB represents the path loss at a reference distance of 1 m and $\alpha_0=2.8$ denotes the path loss exponent. The elevation and azimuth AoAs/AoDs are assumed to be i.i.d. variables following the uniform distribution over $[-\pi/2,\pi/2]$. 
	
	We compare the performance of the proposed designs with the following three baseline schemes.
	\begin{itemize}
		\item[$\bullet$] $\textbf{FA with Fixed Position (FAFP)}$ : The positions of $N_t$ FAs are fixed, and the space of adjacent FAs is $\lambda/2$. The transmitter of BS is equipped with FPA-based uniform planner with $N_t$ antennas, spaced by $\lambda/2$. The dual-functional beamforming is obtained in the same way as Prop.FA .
		\item[$\bullet$] $\textbf{FA with Random Position (FARP)}$: Randomly generate $\tilde{\mathbf{t}}$ satisfying ($\ref{Problema}$), ($\ref{Problemb}$) and ($\ref{Problemc}$). Obtain the sensing SNR in (\ref{sensing_SNR}) by optimizing $\mathbf{W}$ with given $\tilde{\mathbf{t}}$.
		\item[$\bullet$] $\textbf{Alternating position selection (APS)}$: The moving region is quantized into discrete locations with equal-distance $\lambda/2$. The greedy algorithm is employed for the selection of the antenna positions \cite{M13}. Specifically, we select the position of each antenna in order and leave the remaining $N_t -1$ antenna unchanged. For all possible positions of each antenna, we select the position that can maximize the sensing SNR.   
	\end{itemize}

	\subsection{Perfect CSI Case}
	In this subsection, we validate the effectiveness of the proposed design and the baseline schemes under perfect CSI of the communication channel and the sensing channel.
	
	\begin{figure}[t!]
		\centering{\includegraphics[width=0.95\linewidth]{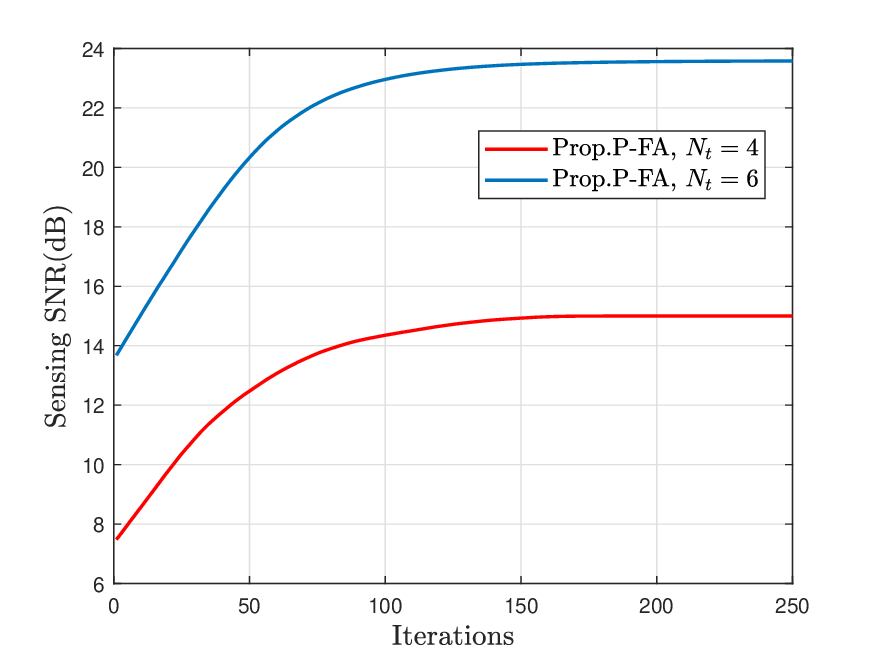}}
		\caption{ Convergence of Algorithm 1.}
		\label{fig1}
	\end{figure}
	
	\subsubsection{Convergence of Algorithm 1} Fig. $\ref{fig1}$ shows the convergence of Algorithm 1 with different numbers of FAs, i.e., $N_t$. We set $A=2\lambda$ under this setup. We can observe that for different $N_t$, the sensing SNRs at the $150$-th iteration are $99\%$ of those at the $200$-th iteration. Thus, in the rest of the subsection, we show the results of Algorithm 1 with $150$ iterations. Then, in the case of $N_t=4$ and $N_t=6$, the converged sensing SNRs increase by 100.5$\%$ and 72.4$\%$ compared to the sensing SNRs of the initial points, respectively.
	
	\begin{figure}[t!]
		\centering{\includegraphics[width=0.95\linewidth]{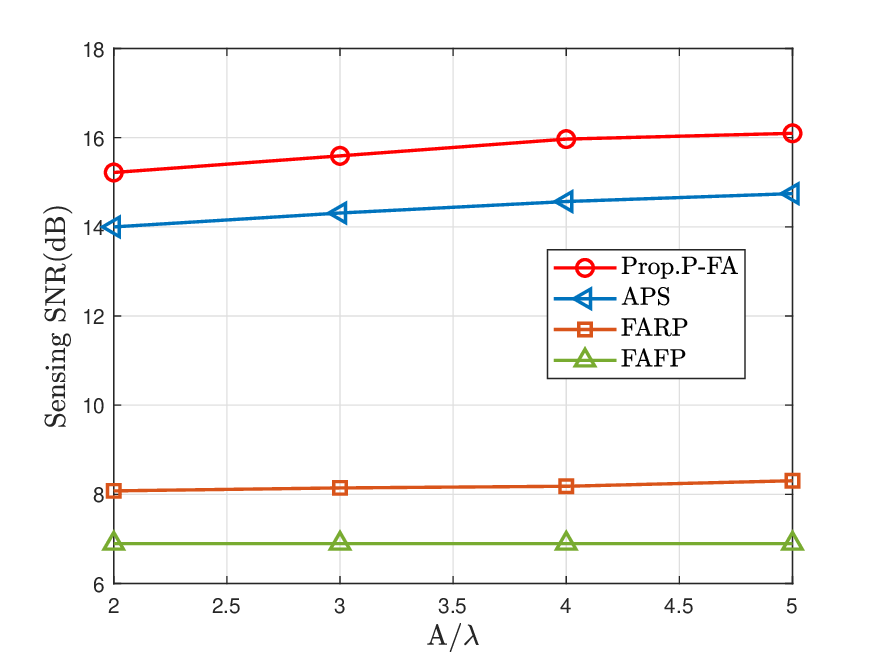}}
		\caption{ Sensing SNR versus the normalized region size.}
		\label{fig2}
	\end{figure}
	
	\subsubsection{Effect of Moving Region Size} Fig. $\ref{fig2}$ illustrates the sensing SNR versus the normalized region size $A/\lambda$ for the Prop.IP-FA and baseline schemes under the setup $\gamma=10dB$. From Fig. $\ref{fig2}$, we have the following observations. Firstly, the sensing SNRs of all designs except FAFP increase with $A/\lambda$, since the increase of the size of the moving region results in a large DoF for the exploration of FAs' spatial domain. Secondly, the proposed design outperforms the three baseline schemes regarding sensing SNR due to the optimization of FAs' positions. Finally, when $A/\lambda$ is greater than 5, the proposed design converges, which indicates that the maximum sensing SNR of the FA-enabled ISAC system can be achieved within a finite transmit region.
	
	\begin{figure}[t!]
		\centering{\includegraphics[width=0.95\linewidth]{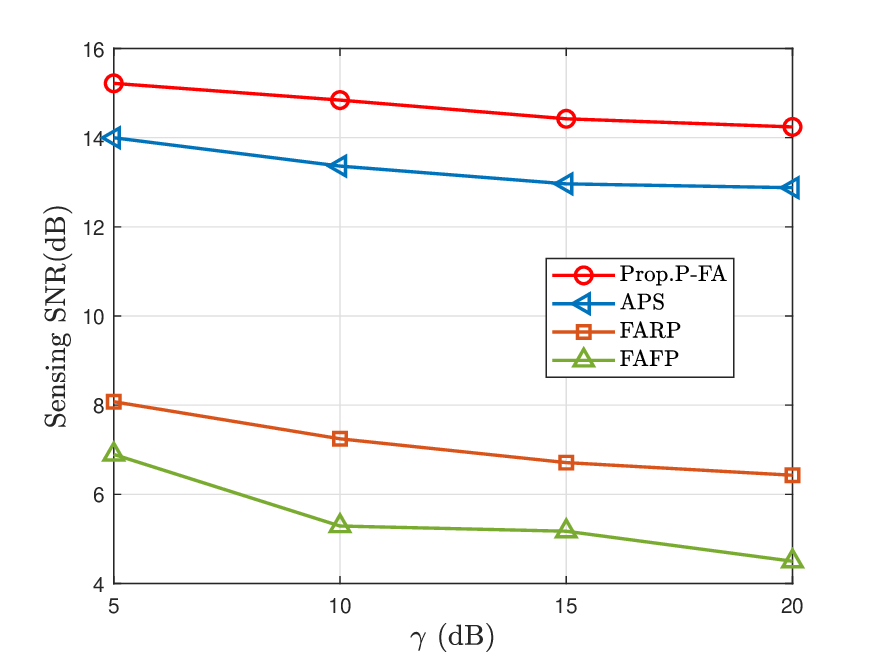}}
		\caption{ Sensing SNR versus the minimum SINR requirement.}
		\label{fig3}
	\end{figure}
	
	\subsubsection{Effect of Minimum SINR Requirement} Fig. $\ref{fig3}$ shows the sensing SNR versus the required threshold SINR of the CU, i.e., $\gamma$, under the setup $A=2\lambda$. From Fig. $\ref{fig3}$, we have the following conclusions. First, the sensing SNRs decrease with $\gamma$ due to the fact that the larger $\gamma$ is, the less energy the BS provides to sense the point target. Moreover, the proposed design outperforms the APS, FAPR, and FAFP by up to $8.72\%-11.26\%$, $88.4\%-121.53$ and $120.7\%-178.85\%$ in terms of sensing SNR, respectively.
	
	\begin{figure}[t!]
		\centering{\includegraphics[width=0.95\linewidth]{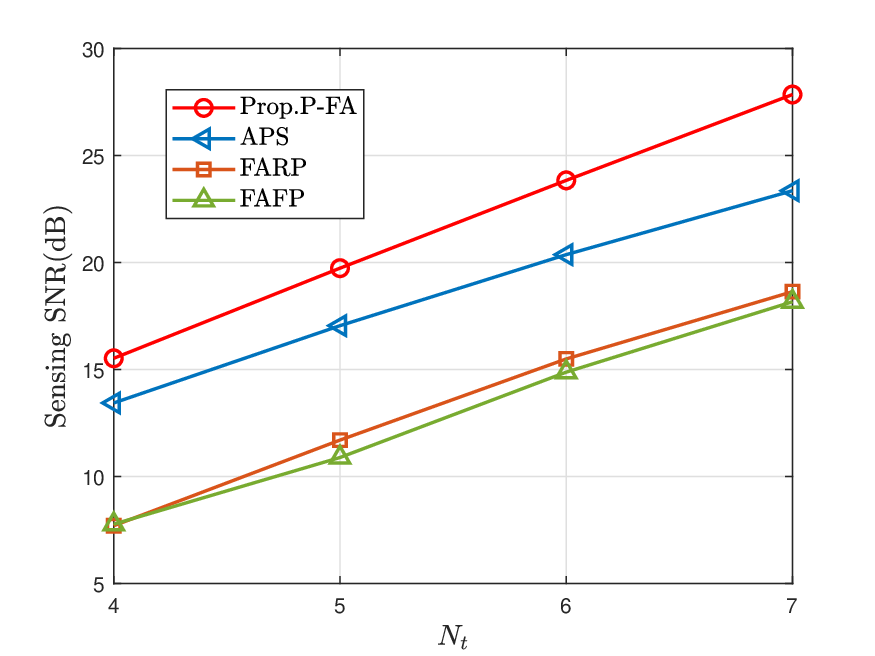}}
		\caption{ Sensing SNR versus the number of fluid antennas.}
		\label{fig4}
	\end{figure}

	\subsubsection{Effect of Number of FAs} Fig. $\ref{fig4}$ plots the sensing SNR versus the number of FAs $N_t$. From Fig. $\ref{fig4}$, the following conclusions can be obtained. First, the sensing SNRs of all designs monotonically increase with $N_t$. Since the more antennas there are, the more spatial diversity gains are obtained. Secondly, the proposed design outperforms other the three baseline schemes due to the full utilization of the spatial resources for FAs. Thirdly, the gaps of sensing SNR between the proposed design and the baseline schemes increase with $N_t$, indicating that the sensing SNR prefers a large $N_t$.

	\begin{figure}[t!]
		\centering{\includegraphics[width=0.95\linewidth]{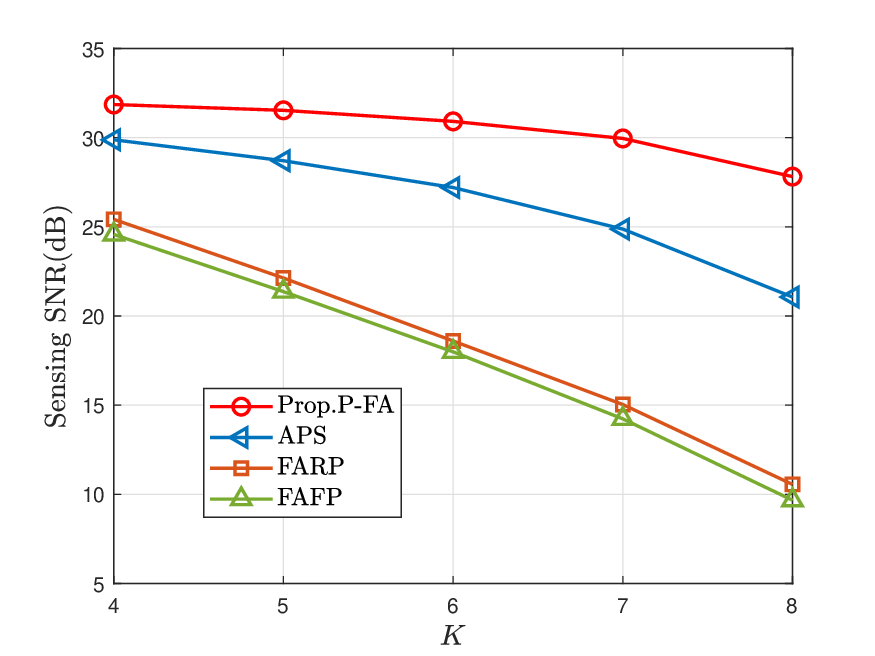}}
		\caption{ Sensing SNR versus the number of Users.}
		\label{fig5}
	\end{figure}
	
	\subsubsection{Effect of Number of Users} In Fig. $\ref{fig5}$, the sensing SNRs versus the number of users $K$, are studied under the setup $N_t = 8$. From Fig. $\ref{fig5}$, we have the following conclusions. Firstly, the sensing SNRs of all designs decrease with $K$ since the more CUs there are, the higher power is demanded to guarantee the QoS of the CUs. It indicates a trade-off in power allocation between the communication SINR and the sensing SNR on the ISAC systems. \mcone{Secondly, the gap of sensing SNR between Prop.P-FA and APS increases with $K$. Since the smaller $K$ is, the more power is used to improve the sensing SNR.}

	\subsection{Imperfect CSI case} \label{convergence_alg2}
	In this subsection, we numerically validate the effectiveness of the proposed design and baseline schemes under imperfect CSI of the communication channel and sensing channel. The estimation errors of azimuth and elevation angles between the BS and the point target are set to $\Phi_{\vartheta}=[40^{\circ},50^{\circ}]$ and $\Phi_{\varphi}=[-35^{\circ},-25^{\circ}]$, respectively. For communication CSI error, we adopt the error percentage $\hat{\epsilon}_k$ for error evaluation, where $\hat{\varepsilon}_k \triangleq \frac{\varepsilon_k}{\Vert \widehat{\mathbf{h}}_k(\tilde{\mathbf{t}}) \Vert}$, $\forall k$. For convenience, we set the values of CSI errors of all CUs to $\varepsilon$.
	
	\begin{figure}[t!]
		\centering
		\subfigure[]{\includegraphics[width=0.95\linewidth]{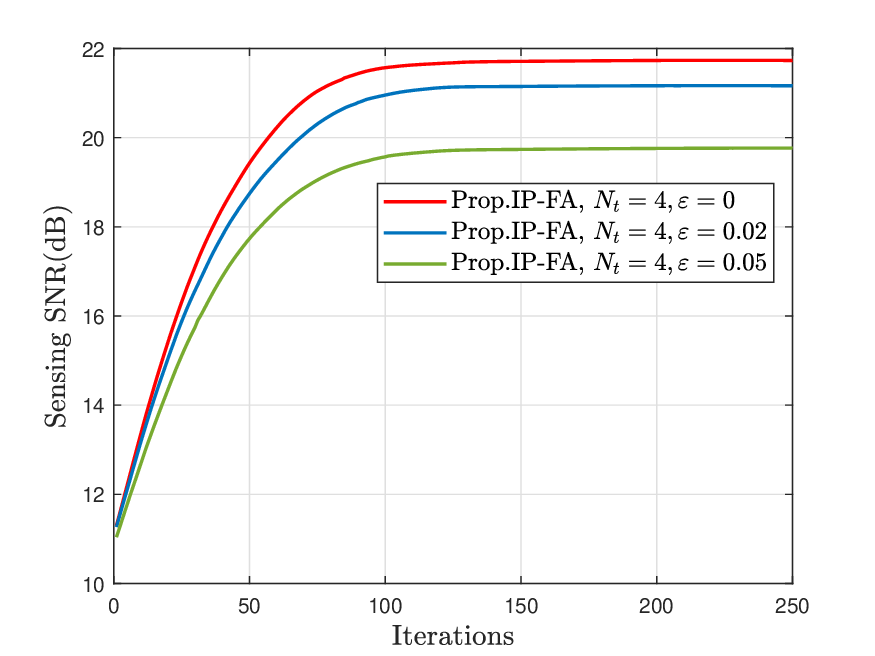}}
		\hspace{10pt}
		\subfigure[]{\includegraphics[width=0.95\linewidth]{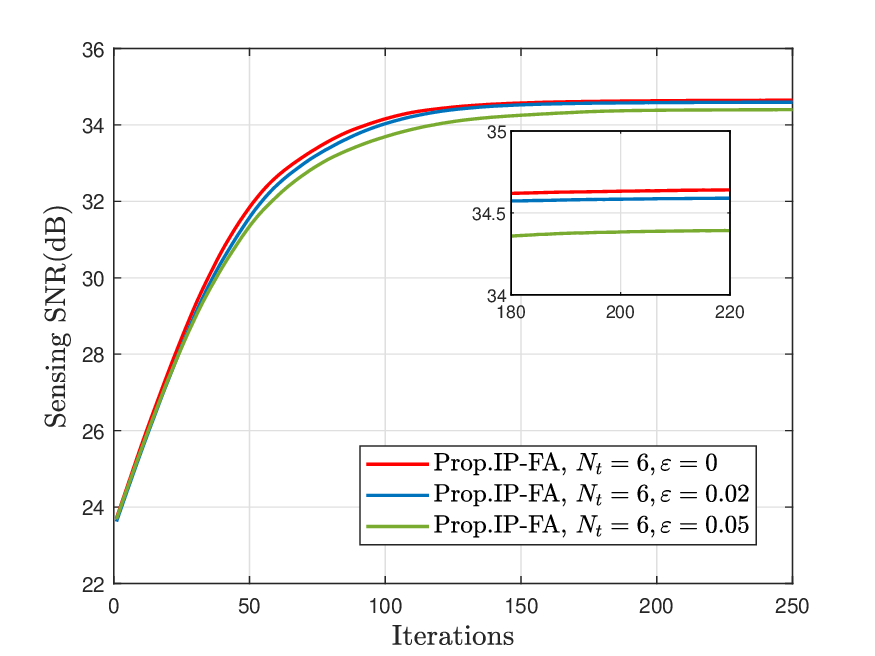}}
		\caption{ Convergence of Algorithm 2.}
		\label{fig6}
	\end{figure}
	
	\subsubsection{Convergence of Algorithm 2} Fig. $\ref{fig6}$ shows the convergence of Algorithm 2 with different CSI error $\varepsilon$. First, we can observe that for different $N_t$ and $\varepsilon$, the sensing SNRs monotonically increase with iteration numbers and converge to stable values. Secondly, \mcone{when $\varepsilon$ is fixed, a large $N_t$ results in a large sensing SNR }since more spatial DoF is exploited to enhance the \mcone{sensing and communication} performance. Thirdly, the sensing SNRs decrease with $\varepsilon$, since the larger CSI error $\varepsilon$ is, the higher power is provided to meet the communication QoS. 
	
	\begin{figure}[t!]
		\centering
		\centering{\includegraphics[width=0.95\linewidth]{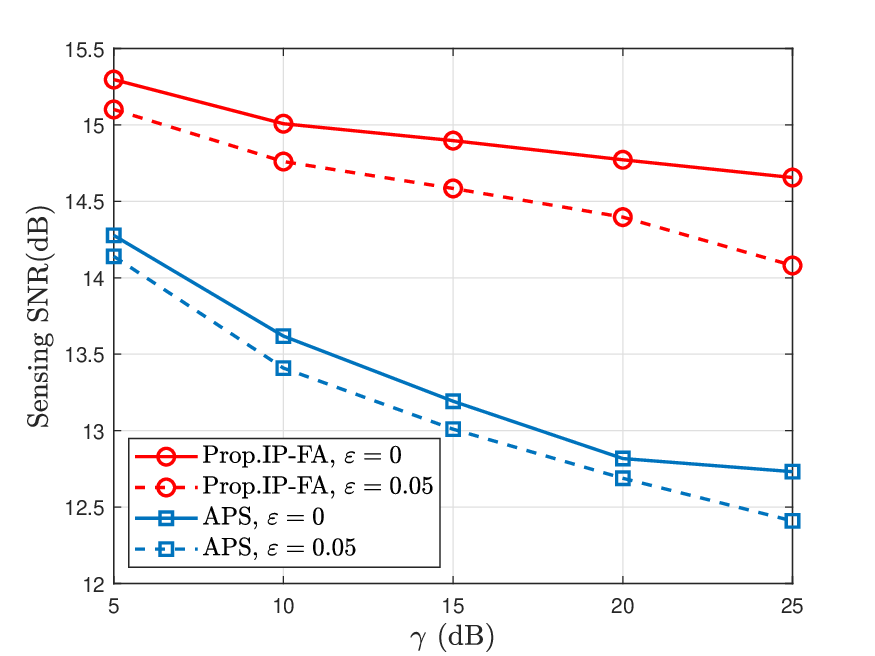}}
		\caption{ Sensing SNR versus Communication SINR.}
		\label{fig7}
	\end{figure}
	
	\subsubsection{Effect of Minimum SINR Requirement} Fig. $\ref{fig7}$ shows the sensing SNRs of the proposed design Prop.IP-FA and the baseline scheme APS versus the required communication SINR $\gamma$. We set $A=2\lambda$ and $\varepsilon=0.05$ or $\varepsilon=0$. From Fig. $\ref{fig7}$, the following conclusions can be obtained. First, the sensing SNRs of Prop.IP-FA and APS decrease with $\gamma$, which also verifies the trade-off on power allocation between the communication SINR and the sensing SNR. Secondly, for fixed $\varepsilon$, Prop.IP-FA outperforms APS due to the robust optimization. Thirdly, the degradation of sensing SNR of Prop.IP-FA caused by the imperfect CSI is small, indicating that the proposed design under imperfect CSI is effective.

	\section{Conclusion}
	
	In this paper, we considered an FA-enhanced ISAC system under perfect and imperfect CSI cases, in which the BS was equipped with multiple FAs and FPAs to simultaneously perform the multi-user communication and the target sensing. Specifically, in two cases, the positions of FAs and the dual-functional beamforming were jointly optimized to maximize the sensing SNR while satisfying the constraints of the finite moving region of FAs, the minimum FA distance, and the minimum SINR requirement per user. First, for the ideal case with perfect CSI of the communication channel and the sensing channel, we proposed an AO-based iterative algorithm in which the dual-functional beamforming and the FA positions were obtained via SDR and SCA techniques. Secondly, for the practical case where the BS only has imperfect CSI of the communication channel and sensing channel, we proposed a new AO-based iterative algorithm with $\mathcal{S}-$Procedure and SCA to solve the formulated problem. Then, the convergence of both proposed algorithms was proved analytically and numerically in the simulation. Numerical results demonstrated that the proposed algorithms can effectively improve the sensing SNR while guaranteeing the communication QoS in both cases.
	\begin{figure*}[ht!]
		\centering
		\begin{equation}
			f_k(\tilde{\mathbf{t}}) = \sum_{i=1}^{N_t}\sum_{l=1}^{L_k}|\sigma_{l,k}|^2\mathbf{R}_k(i,i)
			+\sum_{i=1}^{N_t}\sum_{l=1}^{L_k-1}\sum_{p\neq l}^{L_k}2\mu_{k,i,i,l,p}\cos(\kappa_{k,i,i,l,p})
			+\sum_{i=1}^{N_t-1}\sum_{j=i+1}^{N_t}\sum_{l=1}^{N_t}\sum_{p=1}^{N_t}\mu_{k,i,j,l,p}\cos(\kappa_{k,i,j,l,p}), \label{long_eq}
		\end{equation}	
		\hrulefill
		\vspace*{4pt}
	\end{figure*}
	\begin{figure*}[ht!]
		\normalsize
		\begin{align}
			&\frac{\partial f_k(\tilde{\mathbf{t}})}{\partial x_i} = -\frac{4\pi}{\lambda}\sum_{l=1}^{L_k-1}\sum_{p=1+1}^{L_k}\mu_{k,i,i,l,p}\left(\cos\theta_{k,l}\sin\phi_{k,l}-\cos\theta_{k,p}\sin\phi_{k,p}\right)\sin(\kappa_{k,i,i,l,p}) \nonumber\\ 
			&-\frac{4\pi}{\lambda}\sum_{j=i+1}^{N_t}\sum_{l=1}^{L_k}\sum_{p=1}^{L_k}\mu_{k,i,j,l,p}\cos\theta_{k,l}\sin\phi_{k,l}\sin(\kappa_{k,i,i,l,p})+ \frac{4\pi}{\lambda}\sum_{j=1}^{i-1}\sum_{l=1}^{L_k}\sum_{p=1}^{L_k}\mu_{k,i,j,l,p}\cos\theta_{k,p}\sin\phi_{k,p}\sin(\kappa_{k,i,i,l,p}), \nonumber\\
			&\frac{\partial f_k(\tilde{\mathbf{t}})}{\partial y_i} = -\frac{4\pi}{\lambda}\sum_{l=1}^{L_k-1}\sum_{p=1+1}^{L_k}\mu_{k,i,i,l,p}\left(\sin\theta_{k,l}-\sin\theta_{k,p}\right)\sin(\kappa_{k,i,i,l,p}) -\frac{4\pi}{\lambda}\sum_{j=i+1}^{N_t}\sum_{l=1}^{L_k}\sum_{p=1}^{L_k}\mu_{k,i,j,l,p}\sin\theta_{k,l}\sin(\kappa_{k,i,i,l,p})\nonumber\\
			&+\frac{4\pi}{\lambda}\sum_{j=1}^{i-1}\sum_{l=1}^{L_k}\sum_{p=1}^{L_k}\mu_{k,i,j,l,p}\sin\theta_{k,p}\sin(\kappa_{k,i,i,l,p}), \label{par_f} \\
			&\frac{\partial^2 f_k(\tilde{\mathbf{t}})}{\partial u_m \partial v_m}=-\frac{8\pi^2}{\lambda^2}\sum_{l=1}^{L_k-1}\sum_{p=1+1}^{L_k}\mu_{k,i,i,l,p}\Psi(u)\Psi(v)\sin(\kappa_{k,i,i,l,p})-\frac{8\pi^2}{\lambda^2}\sum_{j=i+1}^{N_t}\sum_{l=1}^{L_k}\sum_{p=1}^{L_k}\mu_{k,i,j,l,p}\psi(u)\psi(v)\sin(\kappa_{k,i,i,l,p})\nonumber\\
			&+\frac{8\pi^2}{\lambda^2}\sum_{j=1}^{i-1}\sum_{l=1}^{L_k}\sum_{p=1}^{L_k}\mu_{k,i,j,l,p}\psi(u)\psi(v)\sin(\kappa_{k,i,i,l,p}), \nonumber\\
			&\frac{\partial^2 f_k(\tilde{\mathbf{t}})}{\partial u_m \partial v_n}=-\frac{8\pi^2}{\lambda^2}\sum_{j=i+1}^{N_t}\sum_{l=1}^{L_k}\sum_{p=1}^{L_k}\mu_{k,i,j,l,p}\psi(u)\psi(v)\sin(\kappa_{k,i,i,l,p})+\frac{8\pi^2}{\lambda^2}\sum_{j=1}^{i-1}\sum_{l=1}^{L_k}\sum_{p=1}^{L_k}\mu_{k,i,j,l,p}\psi(u)\psi(v)\sin(\kappa_{k,i,i,l,p}). \label{par_2_f}
		\end{align}	
		\hrulefill
		\vspace*{4pt}
	\end{figure*}
	\appendix
	\subsection{Derivations of Objective Function of Problem ($\ref{Problem2}$) }
	\subsubsection{Derivations of $\nabla g(\tilde{\mathbf{t}})$ and $\nabla^2 g(\tilde{\mathbf{t}})$ in (\ref{obj_ste})}
	Recall that $\mathbf{Q}=\{q_{i,j}=|q_{i,j}|e^{j\angle q_{i,j}},1\leq i,j\leq N_t\}$, $\tilde{\mathbf{t}}=[\mathbf{t}_1^T,...,\mathbf{t}_{N_t}^T]^T$ and $\mathbf{t}_m=[x_m,y_m]^T$, the gradient vector and Hessian matrix of $g(\tilde{\mathbf{t}})$ w.r.t. $\tilde{\mathbf{t}}$ can be written as:
	\begin{equation}
		\nabla g(\tilde{\mathbf{t}}) = \left[ \frac{\partial g(\tilde{\mathbf{t}})}{\partial x_1}, \frac{\partial g(\tilde{\mathbf{t}})}{\partial y_1},..., \frac{\partial g(\tilde{\mathbf{t}})}{\partial x_{N_t}}, \frac{\partial g(\tilde{\mathbf{t}})}{\partial y_{N_t}} \right]^T,\label{delta_g}
	\end{equation}
	\begin{equation}
		\nabla^2 g(\tilde{\mathbf{t}}) =	\begin{bmatrix}
			\begin{matrix}
				\frac{\partial^2 g(\tilde{\mathbf{t}})}{\partial x_1^2} & \frac{\partial^2 g(\tilde{\mathbf{t}})}{\partial x_1 \partial y_1} \\
				\frac{\partial^2 g(\tilde{\mathbf{t}})}{\partial y_1 \partial x_1} & \frac{\partial^2 g(\tilde{\mathbf{t}})}{\partial y_1^2}
			\end{matrix}
			& \cdots & 
			\begin{matrix}
				\frac{\partial^2 g(\tilde{\mathbf{t}})}{\partial x_1 \partial y_{N_t}} \\ \frac{\partial^2 g(\tilde{\mathbf{t}})}{\partial y_1 \partial y_{N_t}}
			\end{matrix} \\
			\vdots & \ddots & \vdots \\
			\begin{matrix}
				\frac{\partial^2 g(\tilde{\mathbf{t}})}{\partial y_{N_t} \partial x_1} & \frac{\partial^2 g(\tilde{\mathbf{t}})}{\partial y_{N_t} \partial y_1}
			\end{matrix}
			& \cdots &
			\frac{\partial^2 g(\tilde{\mathbf{t}})}{\partial y_{N_t}^2}
		\end{bmatrix}.
		\label{delta_2_g}
	\end{equation}
	For simplicity, we define $\kappa_{m,n}=\rho(\mathbf{t}_m-\mathbf{t}_n,\vartheta,\varphi)$. Combing the above, we have
	\begin{equation}
		\frac{\partial g(\tilde{\mathbf{t}})}{\partial x_m} = -\frac{4\pi}{\lambda}\cos\vartheta \sin\varphi \sum_{n\neq m}^{N_t}|q_{m,n}|\sin(\frac{2\pi}{\lambda}\kappa_{m,n}+\angle q_{m,n}), \nonumber 
	\end{equation}
	\begin{equation}
		\frac{\partial g(\tilde{\mathbf{t}})}{\partial y_m} = -\frac{4\pi}{\lambda}\sin\vartheta \sum_{n\neq m}^{N_t}|q_{m,n}|\sin(\frac{2\pi}{\lambda}\kappa_{m,n}+\angle q_{m,n}), \label{pfk}
	\end{equation} 
	\begin{equation}
		\frac{\partial^2 g(\tilde{\mathbf{t}})}{\partial u_m v_m} = -\frac{8\pi^2}{\lambda}\varphi(u)\varphi(v) \sum_{n\neq m}^{N_t}|q_{m,n}|\cos(\frac{2\pi}{\lambda}\kappa_{m,n}+\angle q_{m,n}),\nonumber
	\end{equation}
	\begin{equation}		
		\frac{\partial^2 g(\tilde{\mathbf{t}})}{\partial u_m v_n} = \frac{8\pi^2}{\lambda}\varphi(u)\varphi(v)|q_{m,n}|\cos(\frac{2\pi}{\lambda}\kappa_{m,n}+\angle q_{m,n}),  \label{p2fk}
	\end{equation}
	where $1\leq m\neq n\leq N_t$, $u,v\in\{x,y\}, \varphi(x)=\cos\vartheta \sin\varphi, \varphi(y)=\sin\vartheta$. 
	\subsubsection{Derivation of $\delta$ in (\ref{obj_ste})}
	Note that $\Vert \nabla^2 g(\tilde{\mathbf{t}}) \Vert_2^2 \leq \Vert \nabla^2 g(\tilde{\mathbf{t}}) \Vert_F^2$ and $\Vert \nabla^2 g(\tilde{\mathbf{t}}) \Vert_2\mathbf{I}_{2N_t} \succeq \nabla^2 g(\tilde{\mathbf{t}})$. Depending on the derivation of $\nabla^2 g(\tilde{\mathbf{t}})$ in ($\ref{delta_2_g}$), $\delta$ is given by:		
	\begin{equation}
			\delta = \frac{16N_t \pi^2 N_t\sqrt{N_t-1}\epsilon}{\lambda^2}, \label{delta}
	\end{equation}
	where $\epsilon = \text{max}(|q_{i,j}|)$.
	
	\subsection{Derivations of $f_k(\tilde{\mathbf{t}})$ in (\ref{f_k})}
	\subsubsection{Derivations of $\nabla f_k(\tilde{\mathbf{t}})$ and $\nabla^2 f_k(\tilde{\mathbf{t}})$ in (\ref{C4_bound})}
	For convenience of expression, we define
	\begin{equation}
		\kappa_{k,i,j,l,p}\triangleq\frac{2\pi}{\lambda}(\rho_{k,l}(\mathbf{t}_i)-\rho_{k,p}(\mathbf{t}_j))+\angle \sigma_{l,k} -\angle \sigma_{p,k}+ \angle \mathbf{R}_k(i,j), \label{kappa}
	\end{equation}
	\begin{equation}
		\mu_{k,i,j,l,p}\triangleq|\sigma_{l,k}||\sigma_{p,k}||\mathbf{R}_k(i,j)|.\label{mu}
	\end{equation}
	Thus, $f_k(\tilde{\mathbf{t}})$ can be transformed into ($\ref{long_eq}$). The gradient vector and Hessian matrix of $f(\tilde{\mathbf{t}})$ w.r.t. $\tilde{\mathbf{t}}$ are given by
	\begin{equation}
	\nabla f_k(\tilde{\mathbf{t}}) = \left[ \frac{\partial f_k(\tilde{\mathbf{t}})}{\partial x_1}, \frac{\partial f_k(\tilde{\mathbf{t}})}{\partial y_1},..., \frac{\partial f_k(\tilde{\mathbf{t}})}{\partial x_{N_t}}, \frac{\partial f_k(\tilde{\mathbf{t}})}{\partial y_{N_t}} \right]^T, \label{delte_f_k}
	\end{equation}
	\begin{equation}
	\nabla^2 f_k(\tilde{\mathbf{t}}) =	\begin{bmatrix}
		\begin{matrix}
			\frac{\partial^2 f_k(\tilde{\mathbf{t}})}{\partial x_1^2} & \frac{\partial^2 f_k(\tilde{\mathbf{t}})}{\partial x_1 \partial y_1} \\
			\frac{\partial^2 f_k(\tilde{\mathbf{t}})}{\partial y_1 \partial x_1} & \frac{\partial^2 f_k(\tilde{\mathbf{t}})}{\partial y_1^2}
		\end{matrix}
		& \cdots & 
		\begin{matrix}
			\frac{\partial^2 f_k(\tilde{\mathbf{t}})}{\partial x_1 \partial y_{N_t}} \\ \frac{\partial^2 f_k(\tilde{\mathbf{t}})}{\partial y_1 \partial y_{N_t}}
		\end{matrix} \\
		\vdots & \ddots & \vdots \\
		\begin{matrix}
			\frac{\partial^2 f_k(\tilde{\mathbf{t}})}{\partial y_{N_t} \partial x_1} & \frac{\partial^2 f_k(\tilde{\mathbf{t}})}{\partial y_{N_t} \partial y_1}
		\end{matrix}
		& \cdots &
		\frac{\partial^2 f_k(\tilde{\mathbf{t}})}{\partial y_{N_t}^2}
	\end{bmatrix}.
	\end{equation}	
	
	For simplicity of the equations, we define
	\begin{align}
		\begin{cases}
		&\Psi(x)\triangleq \left(\cos\theta_{k,l}\sin\phi_{k,l}-\cos\theta_{k,p}\sin\phi_{k,p}\right),\\
		&\Psi(y)\triangleq \left(\sin\theta_{k,l}-\sin\theta_{k,p}\right),\\
		&\psi(x)\triangleq \cos\theta_{k,l}\sin\phi_{k,l},\\
		&\psi(y)\triangleq \cos\theta_{k,l}\sin\phi_{k,l}.
		\end{cases}
	\end{align} 
	
	Thus, the relative terms in $\nabla f_k(\tilde{\mathbf{t}})$ and $\nabla^2 f_k(\tilde{\mathbf{t}})$ are shown in (\ref{par_f}) and (\ref{par_2_f}) at the top of the next page, where $1\leq m\neq n\leq N_t$, $u,v\in\{x,y\}$. The definitions of $\mu_{k,i,j,l,p}$ and $\kappa_{k,i,j,l,p}$ are given in ($\ref{kappa}$) and ($\ref{mu}$). 
	\subsubsection{Derivation of $\zeta_k$ in (\ref{C4_bound})}
	Similar to the calculation of $\delta$ in (\ref{delta}), $\zeta_k$ can be given by:
	\begin{align}
		&~\zeta_k = \frac{16N_t \pi^2}{\lambda^2} ( \sum_{l=1}^{L_k-1}\sum_{p=l}^{L_k}|\sigma_{l,k}||\sigma_{p,k}|\eta_k  \nonumber \\
		&~+ \sum_{l=1}^{L_k}\sum_{p=1}^{L_k}|\sigma_{l,k}||\sigma_{p,k}|(N_t-1)\eta_k ),\label{zeta}
	\end{align}
	where $\eta_k \triangleq \text{max}\{ |\mathbf{R}_{k}(i,j)|\}$.
	
	\subsection{Proof of Proposition 1}
	Let $g_{lb}^{\left(r\right)}(\mathbf{W},\tilde{\mathbf{t}})$ denote the objective function of the problem ($\ref{Problem6}$). First, in step 3 of Algorithm 1, since the optimal value of the problem ($\ref{Problem}$) for given $\tilde{\mathbf{t}}^{\left(r\right)}$ is obtained, we have $\gamma_s(\mathbf{W}^{\left(r\right)},\tilde{\mathbf{t}}^{\left(r\right)})\leq \gamma_s(\mathbf{W}^{\left(r+1\right)},\tilde{\mathbf{t}}^{\left(r\right)})$. Secondly, in step 4 of Algorithm 1, for given $\mathbf{W}^{\left(r+1\right)}$, we have
		\begin{align}
			\gamma_s(\mathbf{W}^{\left(r+1\right)},\tilde{\mathbf{t}}^{\left(r\right)})&~ \overset{(a)}{=} g_{lb}^{\left(r\right)}(\mathbf{W}^{\left(r+1\right)},\tilde{\mathbf{t}}^{\left(r\right)}) \nonumber \\
			&~\overset{(b)}{\leq}g_{lb}^{\left(r\right)}(\mathbf{W}^{\left(r+1\right)},\tilde{\mathbf{t}}^{\left(r+1\right)}) \nonumber \\
			&~\overset{(c)}{\leq}\gamma_s(\mathbf{W}^{\left(r+1\right)},\tilde{\mathbf{t}}^{\left(r+1\right)}),
		\end{align}
	where ($a$) is due to the fact that the second-order Taylor expansion in ($\ref{obj_ste}$) is tight at $\tilde{\mathbf{t}}^{\left(r\right)}$; ($b$) is due to the fact that $g_{lb}^{\left(r\right)}(\mathbf{W}^{\left(r\right)},\tilde{\mathbf{t}}^{\left(r+1\right)})$ is the optimal value of the problem (\ref{Problem6}) at the $r$-th iteration, the equality holds when choosing $\tilde{\mathbf{t}}^{\left(r+1\right)}=\tilde{\mathbf{t}}^{\left(r\right)}$; and ($c$) is because the objective value of the problem ($\ref{Problem5}$) serves as a lower bound of its original problem ($\ref{Problem2}$) at $\tilde{\mathbf{t}}^{\left(r+1\right)}$. Thus, we have $\gamma_s(\mathbf{W}^{\left(r\right)},\tilde{\mathbf{t}}^{\left(r\right)})\leq \gamma_s(\mathbf{W}^{\left(r+1\right)},\tilde{\mathbf{t}}^{\left(r+1\right)})$, which indicates that the objective function of the problem ($\ref{Problem}$) is non-decreasing with iterations of Algorithm~1. Thus, the convergence of Algorithm~1 can be guaranteed.

	\subsection{Derivation of $w_k(\tilde{\mathbf{t}})$ in (\ref{app})}
	\subsubsection{Derivation of $\nabla w_k(\tilde{\mathbf{t}})$ in (\ref{app_2})}
	To calculate the gradient vector and Hessian matrix of $w_k(\tilde{\mathbf{t}})$ w.r.t. $\tilde{\mathbf{t}}$, we rewrite $w_k(\tilde{\mathbf{t}})$ as
	\begin{equation}
		w_k(\tilde{\mathbf{t}})=2\sum_{m=1}^{N_t}\sum_{n=1}^{N_t}\sum_{l=1}^{L_k}|\boldsymbol{\delta}_k(m)||\mathbf{R}_k(m,n)||\sigma_{l,k}|\cos\left( \varpi_k^{n,m,l} \right),
	\end{equation}
	where $\varpi_k^{n,m,l}\triangleq \frac{2\pi}{\lambda}\rho(\mathbf{t}_n,\theta_{k,l},\phi_{k,l})+\angle\boldsymbol{\delta}_k(m)+\angle\mathbf{R}_k(m,n)+\angle\sigma_{l,k}$. Then, the relative terms in $\nabla w_k(\tilde{\mathbf{t}})$ and $\nabla^2 w_k(\tilde{\mathbf{t}})$ can be gotten by the same way as ($\ref{pfk}$) and ($\ref{p2fk}$). To decouple $\tilde{\mathbf{t}}$ and $\boldsymbol{\delta}_k$, we transform $\nabla w_k(\tilde{\mathbf{t}})$ into the following form
	\begin{align}
		&\nabla w_k(\tilde{\mathbf{t}})=\left[ \boldsymbol{\delta}_k\mathbf{s}(\mathbf{t}_1),\boldsymbol{\delta}_k\mathbf{p}(\mathbf{t}_1),...,\boldsymbol{\delta}_k\mathbf{s}(\mathbf{t}_{N_t}),\boldsymbol{\delta}_k\mathbf{p}(\mathbf{t}_{N_t}) \right]^T\nonumber \\
		&-\left[ \mathbf{s}^H(\mathbf{t}_1)\boldsymbol{\delta}_k^H,\mathbf{p}^H(\mathbf{t}_1)\boldsymbol{\delta}_k^H,...,\mathbf{s}^H(\mathbf{t}_{N_t})\boldsymbol{\delta}_k^H,\mathbf{p}^H(\mathbf{t}_{N_t})\boldsymbol{\delta}_k^H) \right]^T, \label{par_w}
	\end{align}
	where $\mathbf{s}(\mathbf{t}_n)=\left[ s_1(\mathbf{t}_n),...,s_{N_t}(\mathbf{t}_n) \right]^T$, $\mathbf{p}(\mathbf{t}_n)=\left[ p_1(\mathbf{t}_n),...,p_{N_t}(\mathbf{t}_n) \right]^T$, $1\leq n \leq N_t$. The relative terms in $\mathbf{s}(\mathbf{t}_n)$ and $\mathbf{p}(\mathbf{t}_n)$ are given by
	\begin{align}
		&s_m(\mathbf{t}_n) = j\frac{2\pi}{\lambda}\mathbf{R}_k(m,n)\sum_{l=1}^{L_k}\cos\theta_{k,l}\sin\phi_{k,l}\sigma_{l,k}e^{j\frac{2\pi}{\lambda}\rho(\mathbf{t}_n,\theta_{k,l},\phi_{k,l})},\nonumber \\
		&p_m(\mathbf{t}_n) = j\frac{2\pi}{\lambda}\mathbf{R}_k(m,n)\sum_{l=1}^{L_k}\sin\theta_{k,l}\sigma_{l,k}e^{j\frac{2\pi}{\lambda}\rho(\mathbf{t}_n,\theta_{k,l},\phi_{k,l})}.
	\end{align} 
	
	\subsubsection{Derivation of $\chi_k$ in (\ref{app_2})}
	We obtain $\chi_k$ by calculating $\Vert \nabla^2 w_k(\tilde{\mathbf{t}}) \Vert_F$:
	\begin{equation}
		\chi_k = \frac{16N_t \pi^2}{\lambda^2}\sqrt{\varepsilon_k}\varpi_k\sum_{l=1}^{L_k}|\sigma_{l,k}|,
	\end{equation}
	where $\varpi_k \triangleq \text{max}(|\mathbf{R}_k(n,m)|)$.
	
	\subsection{Proof of Proposition 2}
	Let $z(\left\{\mathbf{T}_k\right\},\tilde{\mathbf{t}})$ and $z_{lb}^{\left(j\right)}(\left\{\mathbf{T}_k\right\},\tilde{\mathbf{t}})$ denote the objective values of the problem ($\ref{Problem7}$) and problem ($\ref{Problem12}$). First, in step 3 of Algorithm 2, since the optimal value of the problem ($\ref{Problem10}$) for given $\tilde{\mathbf{t}}^{\left(j\right)}$ is obtained, we have $z(\left\{\mathbf{T}_k\right\}^{\left(j\right)},\tilde{\mathbf{t}}^{\left(j\right)})\leq z(\left\{\mathbf{T}_k\right\}^{\left(j+1\right)},\tilde{\mathbf{t}}^{\left(j\right)})$. Secondly, in step 4 of Algorithm 2, for given $\left\{\mathbf{T}_k\right\}^{\left(j+1\right)}$, we have
	\begin{align}
		z(\left\{\mathbf{T}_k\right\}^{\left(j+1\right)},\tilde{\mathbf{t}}^{\left(j\right)})&~ \overset{(a)}{=} z_{lb}^{\left(j\right)}(\left\{\mathbf{T}_k\right\}^{\left(j+1\right)},\tilde{\mathbf{t}}^{\left(j\right)}) \nonumber \\
		&~\overset{(b)}{\leq}z_{lb}^{\left(j\right)}(\left\{\mathbf{T}_k\right\}^{\left(j+1\right)},\tilde{\mathbf{t}}^{\left(j+1\right)}) \nonumber \\
		&~\overset{(c)}{\leq}z(\left\{\mathbf{T}_k\right\}^{\left(j+1\right)},\tilde{\mathbf{t}}^{\left(j+1\right)}),
	\end{align}
	where ($a$) is due to the fact that the second-order Taylor expansions in ($\ref{p11a_l}$), ($\ref{p11_2}$), and ($\ref{app_2}$), are tight at $\tilde{\mathbf{t}}^{\left(j\right)}$; ($b$) holds since $z_{lb}^{\left(j\right)}(\left\{\mathbf{T}_k\right\}^{\left(j\right)},\tilde{\mathbf{t}}^{\left(j+1\right)})$ is the optimal value of the problem (\ref{Problem12}) at the $j$-th iteration, the equality holds when choosing $\tilde{\mathbf{t}}^{\left(j+1\right)}=\tilde{\mathbf{t}}^{\left(j\right)}$; and ($c$) is because the objective value of the problem ($\ref{Problem12}$) serves as a lower bound of its original problem ($\ref{Problem7}$) at $\tilde{\mathbf{t}}^{\left(j+1\right)}$. Thus, we have $z(\left\{\mathbf{T}_k\right\}^{\left(j\right)},\tilde{\mathbf{t}}^{\left(j\right)})\leq z(\left\{\mathbf{T}_k\right\}^{\left(j+1\right)},\tilde{\mathbf{t}}^{\left(j+1\right)})$, which indicates that the objective function of the problem ($\ref{Problem7}$) is non-decreasing with iterations of Algorithm~2. Thus, the convergence of Algorithm~2 can be guaranteed.

\end{document}